\documentclass[apj]{emulateapj}
\usepackage{natbib,graphicx,longtable,afterpage}

\def\Ha{\text{H$\alpha$}}
\def\Hb{\text{H$\beta$}}
\def\Hg{\text{H$\gamma$}}
\def\MgII{\text{Mg\,\textsc{ii}}}
\def\FeII{\text{Fe\,\textsc{ii}}}
\def\NII{\text{[N\,\textsc{ii}}]}
\def\OI{\text{[O\,\textsc{i}}]}
\def\OII{\text{[O\,\textsc{ii}}]}
\def\OIII{\text{[O\,\textsc{iii}}]}
\def\SII{\text{[S\,\textsc{ii}}]}
\def\SIII{\text{[S\,\textsc{iii}}]}

\shorttitle{Eddington Ratio Distribution of Broad-line AGNs}
\shortauthors{Suh et al.}

\begin{document}

\title{Eddington ratio distribution of X-ray selected Broad-line AGNs at $1.0<z<2.2$\altaffilmark{$\dagger$}}
\altaffiltext{$\dagger$}{Based in part on data collected at Subaru Telescope, which is operated by the National Astronomical Observatory of Japan.}

\author{Hyewon Suh\altaffilmark{1}, G\"unther Hasinger\altaffilmark{1}, Charles Steinhardt\altaffilmark{2,3}, John D. Silverman\altaffilmark{3}, and Malte Schramm\altaffilmark{4,3}}

\affil{$^{1}$ Institute for Astronomy, University of Hawaii, 2680 Woodlawn Drive, Honolulu, HI 96822, USA \\
$^{2}$ California Institute of Technology, 1200 East California Boulevard, Pasadena, CA 91125, USA \\
$^{3}$Kavli Institute for the Physics and Mathematics of the Universe, Todai Institutes for Advanced Study, the University of Tokyo, Kashiwa, Japan 277-8583 (Kavli IPMU, WPI) \\
$^{4}$National Astronomical Observatory of Japan, Mitaka, Tokyo 181-8588, Japan}


\begin{abstract}
We investigate the Eddington ratio distribution of X-ray selected broad-line active galactic nuclei (AGN) in the redshift range $1.0<z<2.2$, where the number density of AGNs peaks. Combining the optical and Subaru/FMOS near-infrared spectroscopy, we estimate black hole masses for broad-line AGNs in the {\it Chandra} Deep Field-South (CDF-S), Extended {\it Chandra} Deep Field-South (E-CDF-S), and the {\it XMM-Newton} Lockman Hole ({\it XMM}-LH) surveys. AGNs with similar black hole masses show a broad range of AGN bolometric luminosities, which are calculated from X-ray luminosities, indicating that the accretion rate of black holes is widely distributed. We find that a substantial fraction of massive black holes accreting significantly below the Eddington limit at $z\lesssim2$, in contrast to what is generally found for luminous AGNs at high redshift. Our analysis of observational selection biases indicates that the ``AGN cosmic downsizing" phenomenon can be simply explained by the strong evolution of the co-moving number density at the bright end of the AGN luminosity function, together with the corresponding selection effects. However, it might need to consider a correlation between the AGN luminosity and the accretion rate of black holes that luminous AGNs have higher Eddington ratios than low-luminosity AGNs in order to understand the relatively small fraction of low-luminosity AGNs with high accretion rates in this epoch. Therefore, the observed downsizing trend could be interpreted as massive black holes with low accretion rates, which are relatively fainter than less massive black holes with efficient accretion.
\end{abstract}



\section{Introduction}

Disentangling the origin and the mass accretion history of black holes is one of the most outstanding issues for understanding the fundamental processes of galaxy formation and evolution. Observations have shown that supermassive black holes are tightly linked with their host galaxies, as revealed by correlations between the black hole mass and the bulge stellar mass, i.e., ${\rm M_{BH}-M_{stellar}}$ relation \citep{Kormendy95, Magorrian98, Gultekin09, Schulze11, McConnell13} and the velocity dispersion, i.e., ${\rm M_{BH}-\sigma}$ relation \citep{Ferrarese00, Gebhardt00, Merritt01, Tremaine02, Gultekin09, Graham11, McConnell13, Woo13}. Furthermore, it has been widely accepted that the growth of active galactic nuclei (AGN) and the star formation history undergo very similar evolutionary behavior through cosmic time, where the peaks of most luminous AGNs and powerful star-forming galaxies occur at a similar cosmic epoch ($z=2-3$) with a dramatic decline towards low redshift, while the moderate-luminosity AGNs and the bulk of star-forming galaxies peak at lower redshift ($z\lesssim1$) (see e.g. \citealt{Shankar09, Madau14}). This seems to imply that the interaction between the nuclear activity and the star formation in galaxies plays a crucial role in the evolution of black holes and galaxies over cosmic time.

The AGN luminosity function and its evolution are key observational properties for understanding the accretion history onto the black holes. Observational studies have revealed a ``cosmic downsizing'' or ``anti-hierarchical'' phenomenon in the black hole growth, which means that the characteristic luminosity of AGNs decreases with time. The co-moving number density of luminous AGNs peaks at higher redshift ($z\sim2$) than moderate-luminosity AGNs, which peaks at $z<1$ \citep{Giacconi02, Cowie03, Steffen03, Ueda03, Barger05, Hasinger05, LaFranca05, Hopkins07, Silverman08a}. This AGN cosmic downsizing trend is seen across a wide range of the electromagnetic spectrum in X-ray, optical, infrared, and radio wavebands \citep{Bongiorno07, Cirasuolo07}. If AGN luminosity would strictly correlate with black hole mass, this finding would imply that more massive black holes formed before lower-mass black holes, which is in apparent contradiction to the currently favored hierarchical structure formation paradigm based on the standard cold dark matter model. In the hierarchical framework, more massive halos grow over time hierarchically via subsequent merging and smooth accretion among low mass halos. 

The AGN cosmic downsizing, however, is observed in luminosity, and thus the downsizing phenomenon can also be interpreted assuming a relationship between the AGN luminosity and the black hole mass as a function of redshift. Black holes are assumed to undergo several episodes of a significant gas accretion with complex hydrodynamic and magnetic processes, along with relativistic effects during which this accretion powers AGNs (e.g. \citealt{Springel05, Choi12}). The most luminous AGNs are interpreted as results of major mergers. A substantial starburst occurs as a result of major mergers, and some of the gas eventually reaches the black hole at the center of a galaxy, triggering the AGN activity (see e.g. \citealt{DiMatteo05}). On the other hand, moderate-luminosity AGNs are suggested to be products of modest accretion, in which case the gas accretion via internal, secular processes trigger the AGN activity (e.g. \citealt{Hopkins06, Fanidakis12}). An AGN with black hole mass of M$_{\rm BH}$ can produce the maximum luminosity via the Eddington limit (L$_{\rm Edd}$) at which the radiation pressure by the accretion of the infalling matter balances the gravitational attraction of the black hole for spherically symmetric time-invariant accretion. By estimating the Eddington ratios, the ratio of the AGN bolometric luminosity and the Eddington luminosity (L$_{\rm bol}$/L$_{\rm Edd}$), it can be determined whether the accretion rate of black holes can change over cosmic time. One might have expected a correlation between black hole masses and AGN bolometric luminosities, but if there is a range in accretion rates and/or efficiencies, the relation will be weaker. Thus, in order to investigate the observed downsizing trend in black hole growth, it is important to explore the efficiency of gas accretion during the active phases of black holes. Therefore, the black hole mass and the bolometric luminosity are the key parameters to understanding the evolutionary picture for AGNs. 

Large, modern photometric and spectroscopic surveys open up a new regime for studying a large sample of AGNs (e.g. Sloan Digital Sky Survey (SDSS), \citealt{Schneider10, Shen11}). Many efforts have been made to describe the properties of thousands of AGNs (e.g. \citealt{McLure04, Vestergaard09, Steinhardt10, Choi14}). In previous studies, the Eddington ratio has been assumed to be close to the Eddington limit regardless of redshift and luminosities. \citet{Marconi04} suggest that the Eddington ratios of local black holes are in the range between 0.1 and 1.7, suggesting that black hole growth takes place during luminous accretion phases close to the Eddington limit at high redshift. \citet{Kollmeier06} present that the AGN population is dominated by narrowly distributed near-Eddington accretion rate objects, with a median of 0.1 and a dispersion of 0.3 dex, also suggesting that supermassive black holes gain most of their mass while radiating close to the Eddington limit. However, it is difficult to draw any conclusions about the underlying distribution of the Eddington ratio because the shallowness of the large wide area surveys imposes severe restrictions on the combinations of AGN luminosities and black hole masses that are observable, especially at $z>1$. Recent studies have shown that there is a wide spread in the range of the Eddington ratios (e.g. \citealt{Babic07, Fabian08, Schulze10, Kelly10}). \citet{Lusso12} find that the Eddington ratio increases with redshift for AGNs at any given black hole masses. They also show that the Eddington ratio increases with AGN bolometric luminosity, while no clear evolution with redshift is seen. A wide range of Eddington ratios indicates that their luminosity is not directly related to the black hole mass. Therefore, it is necessary to consider a wide range of Eddington ratios with respect to the AGN luminosity and the black hole mass in order to understand the accretion growth history of the black holes.

Unfortunately, the detailed follow-up study in the redshift interval $z=1-2$, where the AGN downsizing appears, has been difficult because of the lack of emission line diagnostics in the optical wavelength range, which is often referred to as the redshift desert. The strong Balmer emission lines, \Ha~and \Hb, are redshifted to 13126\AA~and 9722\AA~at z=1, respectively. The advent of the sensitive near-infrared (NIR) spectrograph Fiber Multi Object Spectrograph (FMOS) on the Subaru telescope finally enables us to determine the black hole mass in the key redshift interval $z=1-2$ using the Balmer lines that are the same lines for which the black hole masses are calibrated at low redshift. This redshift range is of particular interest, as it is the epoch that a significant part of the accretion growth of black holes takes place, where the AGN density peaks and where optical spectroscopy cannot easily determine the redshifts and properties of many of the AGNs.

In this paper, we investigate the Eddington ratios for X-ray selected broad-line AGNs in the {\it Chandra} Deep Field-South (CDF-S), Extended {\it Chandra} Deep Field-South (E-CDF-S), and the {\it XMM}-Newton Lockman Hole ({\it XMM}-LH) fields. Absorption-corrected X-ray luminosities together with bolometric corrections will allow an estimate of bolometric luminosities of AGNs. The advantage of using X-ray luminosities to derive AGN bolometric luminosities is that they are relatively less affected by the presence of obscuration and contamination effects from the host galaxy. We determine black hole masses using Subaru/FMOS NIR spectroscopic observations and available optical spectroscopy from the literature. We also investigate the possible biases due to systematics and selection effects on the observed data.

Throughout this paper we assume a $\Lambda$CDM cosmology with $\Omega_{m} = 0.3$, $\Omega_{\Lambda}=0.7$ and H$_{0} = 70$ km s$^{-1}$ Mpc$^{-1}$.


\begin{figure*}
\centering
\includegraphics[width=0.49\textwidth]{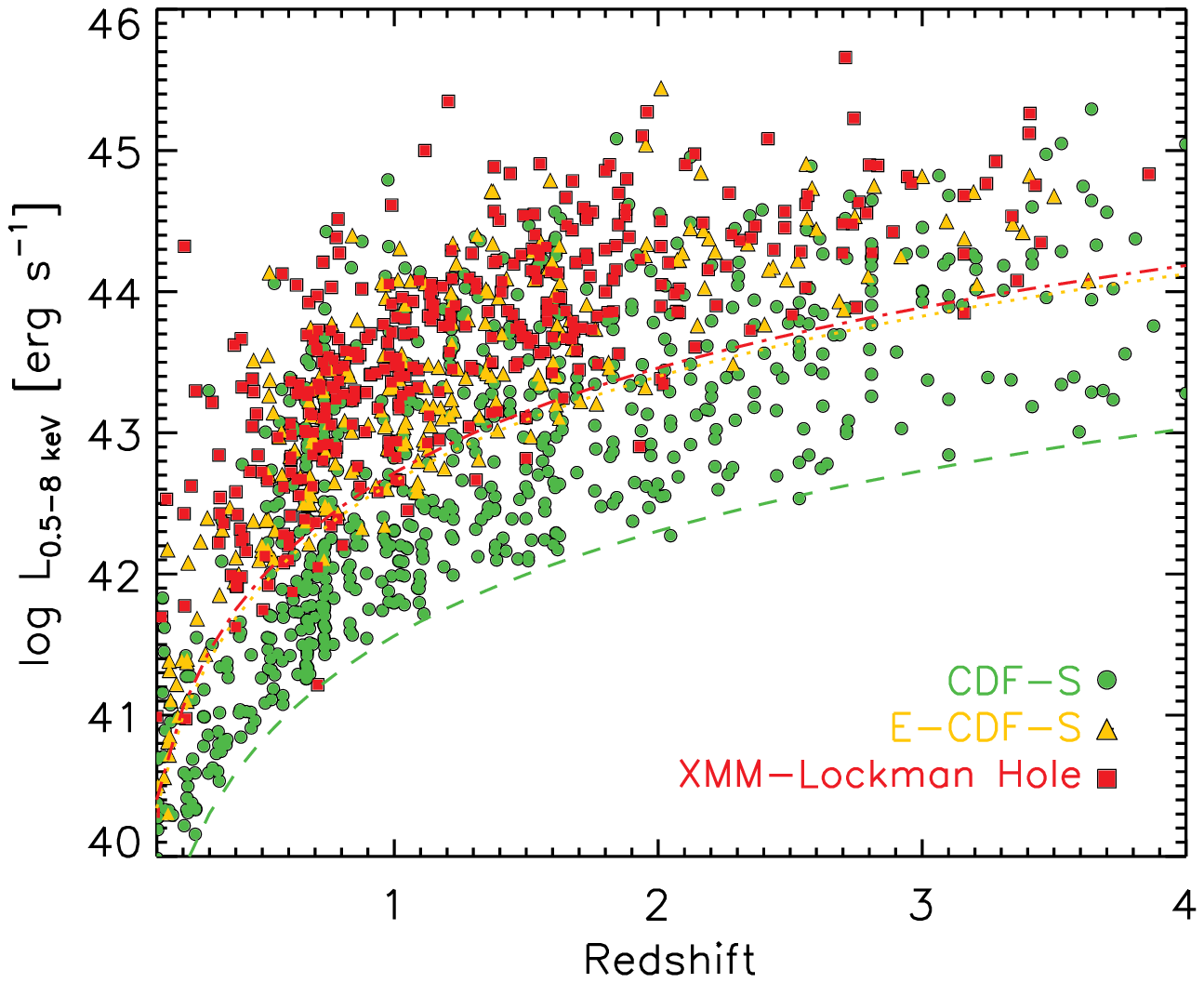}
\includegraphics[width=0.49\textwidth]{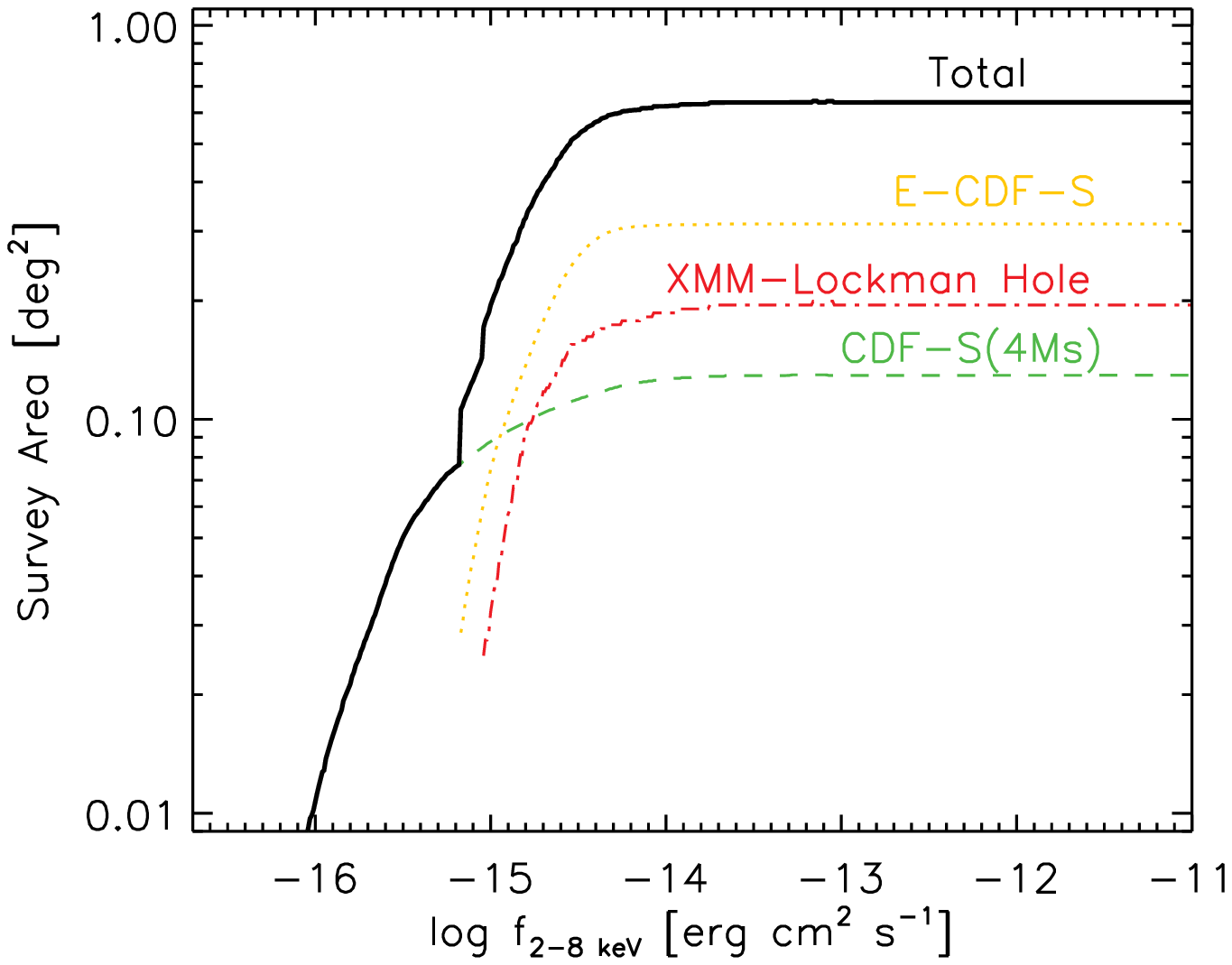}
\caption{{\it Left:} The absorption-corrected 0.5-8 keV X-ray luminosity versus redshift (spectroscopic or photometric) from the X-ray surveys as labeled. {\it Right:} Survey area coverage as a function of X-ray flux. The black line represent the total combined area of all surveys.}
\end{figure*}

\section[]{Sample Selection}

An X-ray survey is practically the most efficient way of finding AGNs over a wide range of luminosities and redshifts. The deep surveys with {\it Chandra} (i.e., {\it Chandra} Deep Field-North and {\it Chandra} Deep Field-South; see \citealt{Brandt05, Brandt10}) and {\it XMM-Newton} (i.e., Lockman Hole; see \citealt{Hasinger01, Rovilos11}) allow us to detect a fair sample of low-luminosity (${\rm 42<log~L_{X}<44}$) AGNs out to $z\sim5$, providing a unique opportunity of studying AGN evolution. Besides, these fields are the best window for the deepest and cleanest images at a variety of wavelengths due to the remarkably low Galactic line-of-sight H\textsc{i} column density (i.e., ${\rm N_{H}=8.8\times10^{19}~cm^{-2}}$ for CDF-S; \citealt{Stark92}, ${\rm N_{H}=5.7\times10^{19}~cm^{-2}}$ for {\it XMM}-LH; \citealt{Lockman86}).

We start by selecting a sample of AGNs based on comprehensive catalogs of X-ray sources observed in the CDF-S, E-CDF-S, and {\it XMM}-LH fields, described below.

\subsection{Chandra Deep Field South}

The catalog for the 4 Ms CDF-S, which is the deepest {\it Chandra} survey covering an area of 464.5 arcmin$^{2}$, contains 740 X-ray sources providing the most sensitive $0.5-8$ keV view of the distant universe \citep{Xue11}. The survey reaches flux limits of ${\rm 3.2\times10^{-17},~9.1\times10^{-18},~and~5.5\times10^{-17}~erg~cm^{-2}~s^{-1}}$ for the full ($0.5-8$ keV), soft ($0.5-2$ keV), and hard ($2-8$ keV) bands, respectively. 674 out of the 740 main-catalog sources have either spectroscopic or photometric redshifts, yielding an overall redshift completeness of $\sim$91\%.

In addition to the 4Ms CDF-S point source catalog, we make use of the E-CDF-S observations that have been analyzed and cataloged by \citet{Lehmer05} and \citet{Silverman10}, providing a sample of 762 distinct X-ray point sources with either spectroscopic or photometric redshifts. Of the 762 E-CDF-S main catalog sources, 523 sources were used since 239 sources were also present in the 4 Ms CDF-S catalog. We have an excellent redshift completeness of $\sim$95\% (498/523). This survey reaches sensitivity limits of ${\rm 1.1\times10^{-16}~and~6.7\times10^{-16}~ergs~cm^{-2}~s^{-1}}$ for the soft ($0.5-2.0$ keV) and hard ($2-8$ keV) bands, respectively. 

\subsection{XMM-Newton Lockman Hole}

The catalog of the 409 {\it XMM}-LH X-ray sources is presented in \citet{Brunner08}, with sensitivity limits of ${\rm 1.9\times10^{-16},~9\times10^{-16},~and~1.8\times10^{-15}~erg~cm^{-2}~s^{-1}}$ in the soft ($0.5-2$ keV), hard ($2-10$ keV), and very hard ($5-10$ keV) bands, respectively. \citet{Fotopoulou12} provide spectroscopic or photometric redshifts for the {\it XMM}-LH X-ray sources. There is a reasonably high redshift completeness with 92\% (376 out of 409). Although the sensitivity limit of the {\it XMM}-LH survey is much higher than that of the {\it Chandra} survey, the larger field of view of {\it XMM}-LH (${\rm 25\times25~arcmin^{2}}$) offers a significant sample of bright AGNs while the CDF-S observation provides the fainter tail of AGNs. 

\subsection{X-ray selected AGN Sample}

We generate a total of 1548 X-ray selected AGNs which have reliable spectroscopic or photometric redshift identifications from the X-ray catalogs containing a total of 1672 X-ray sources. We show the X-ray sources with either spectroscopic or photometric redshift from the X-ray surveys as labeled in Figure 1. The absorption-corrected $0.5-8$ keV X-ray luminosity of AGNs as a function of redshift (spectroscopic or photometric) is shown in the left panel of Figure 1. The luminosity of AGNs in the redshift range $1.0<z<2.2$ is distributed between ${\rm L_{0.5-8~keV} = 10^{42.5}~and~10^{45.5} erg/s}$. In the right panel, we show the sky coverage for the individual surveys and the total sample used. As shown, the total sky area is ${\rm \sim0.7~deg^{2}}$ with the narrow deep CDF-S field and the shallower wide-area E-CDF-S and XMM-LH surveys. The deep CDF-S survey improves the AGN sample at low luminosities, while the E-CDF-S and XMM-LH surveys, of shallower depth but of wider area, effectively supply the more luminous AGNs.


\section[]{Spectroscopic data}

\subsection{Subaru/FMOS Near-Infrared Observations}

We performed NIR spectroscopic observations for the AGN sources with the FMOS \citep{Kimura10} high-resolution spectrographs on the Subaru telescope. FMOS provides up to 400 1.2$''$ diameter fibers in the circular 30$'$ diameter field of view. In the high resolution mode, the FMOS spectral coverage is divided into four bands, which are J-short ($0.92-1.12~\mu$m), J-long ($1.11-1.35~\mu$m), H-short ($1.40-1.60~\mu$m), and H-long ($1.60-1.80~\mu$m) with a spectral resolution of R$=\lambda/\Delta\lambda\sim2200$. The Cross-Beam Switching (CBS) mode, in which two fibers are allocated to each target, was used for optimal sky subtraction of faint sources. The fibers in each pair are separated by 60 arcseconds, alternating between one for the target and the other one simultaneously placed on the sky, so that sky subtraction is not affected by time variation of sky brightness.

The primary targets are X-ray selected AGNs in the CDF-S, E-CDF-S, and {\it XMM}-LH surveys with either spectroscopic or photometric redshifts in the range $1.0<z<2.2$, and J magnitudes brighter than 22.5 mag. The FMOS J-band and H-band observations cover the \Ha~and/or \Hb~lines in the redshift range $z=0.7-2.7$. The data was obtained during 2012-2013, shown in Table 1. We observed for a total integration time of 3.5$-$4 hours while accumulating 28-30 frames with an exposure time of 900 seconds per frame. The weather conditions were acceptable, with seeing typically in the range of $0.''6$ to $1.''2$.

\begin{deluxetable}{lccc}
\tabletypesize{\scriptsize}
\tablewidth{0pt}
\tablecaption{SUBARU FMOS Spectroscopic Observations}
\tablehead{
   \colhead{\textbf{Date}} &
   \colhead{\textbf{Field}} &
   \colhead{\textbf{Spectrograph}}}
\startdata
2012 Mar 25 & XMM-LH & J-long \\
2012 Mar 26 & XMM-LH & H-long \\
2012 Dec 28 & CDF-S & J-long \\
2012 Dec 29 & CDF-S, XMM-LH & H-short \\
2012 Dec 30 & CDF-S, XMM-LH & H-long \\
2013 Jan 19 & CDF-S & H-long \\
2013 Jan 20 & CDF-S & J-long \\
2013 Jan 21 & CDF-S & J-long \\
2013 Feb 24 & XMM-LH & H-short \\
2013 Oct 23 & CDF-S & J-long \\
2013 Oct 24 & CDF-S & H-long  
\enddata
\end{deluxetable}

We reduced the data using the publicly available software FIBRE-pac (FMOS Image-Based REduction package; \citealt{Iwamuro12}), which is an IRAF-based reduction tool for FMOS. This procedure includes background subtraction, corrections of detector cross talk, bias difference, bad pixels, the spectral distortion, and the removal of residual airglow lines. Individual frames were combined into an ensemble image, and wavelength and flux calibration were carried out. For the absolute flux calibration, the bright (J$_{\rm AB}$=15-18 mag) stars in each frame were used as a spectral reference. The flux of the reference star was estimated and compared with the photometric data in the catalog. All the spectra were divided by the reference spectrum, and then multiplied by the expected spectrum of the reference star. Apart from the calibration of slit losses through the spectroscopic reference star we do not apply further calibration corrections for our sample of AGNs, since we assume that the reference star corrects most of the slit losses for the point-like sources. While systematic effects like weather conditions, position accuracy still may cause differential flux losses across the field of view, the effect of these systematic errors on black hole masses should be small, since the black hole mass is a function of the square root of the luminosity (see Section 5.2). Finally, the one-dimensional spectrum of each object was extracted from the calibrated image, together with the associated noise spectra.

With the fully reduced 1- and 2-dimensional spectra, we determined the redshift through the identification of prominent emission line features. Each spectrum was visually inspected by Suh and Hasinger individually using the SpecPro \citep{Masters11} environment, which is an IDL-based interactive program for viewing and analyzing spectra. We assigned a quality flag to each redshift to indicate the reliability of the redshift determination. Altogether 825 X-ray sources were observed in the combined CDF-S, E-CDF-S, and {\it XMM}-LH fields, of which 262 sources are spectroscopically identified. It is noteworthy that we identified new spectroscopic redshifts for 135 X-ray selected AGNs. The redshift identifications are summarized in Appendix Table 2. 

\subsection{Optical spectroscopy}

In addition to NIR spectra, we use existing optical spectroscopy that includes a detection of a broad \MgII~emission line, shown to be a reliable probe of black hole mass at $z>1$ (e.g. \citealt{McLure02, Shen12, Matsuoka13}). Optical spectroscopy has been obtained in the CDF-S, E-CDF-S, and {\it XMM}-LH fields \citep{Lehmann00, Lehmann01, Szokoly04, Silverman10, Barger14}, providing spectroscopic redshifts for X-ray sources. \citet{Szokoly04} present the results of spectroscopic follow-up for the CDF-S, which were observed at the VLT with the FORS1/FORS2 spectrographs for {\it Chandra} sources. Furthermore, \citet{Silverman10} provide high-quality optical spectra in the E-CDF-S. 283 {\it Chandra} sources are observed with deep exposures (2-9 hr per pointing) using multi-slit facilities on both VLT/VIMOS and Keck/DEIMOS. \citet{Lehmann00, Lehmann01} offer spectroscopy of the ROSAT Deep Surveys in the Lockman Hole using low-resolution Keck spectra. We compile the existing optical observations of X-ray AGNs from these deep spectroscopic surveys.


\section{AGN Bolometric Luminosity}

The bolometric luminosity of AGNs can be estimated from the X-ray luminosity by applying a suitable bolometric correction. In order to estimate an accurate total intrinsic luminosity radiated by the AGN accretion disc, it is necessary to constrain the absorption-corrected intrinsic X-ray luminosity because it is often obscured and also includes reprocessed radiation. We thus derive the absorption corrected rest-frame X-ray luminosity and determine the bolometric luminosity with the bolometric correction. To account for the dependence of the optical to X-ray flux ratio $\alpha_{\rm ox}$ on luminosity, we use the luminosity-dependent bolometric correction factor (see e.g. \citealt{Vignali03, Marconi04, Hopkins07, Lusso12}). Despite some difference between the luminosity-dependent bolometric correction factor among different studies (e.g. \citealt{Lusso12} predicted lower bolometric correction at high bolometric luminosity with respect to that predicted by \citealt{Marconi04} and \citealt{Hopkins07}), the same trend of increasing bolometric correction at increasing bolometric luminosity is observed within the scatter.

We compute the intrinsic X-ray luminosity of broad-line AGNs following \citet{Xue11}. As a first step, we assume the intrinsic X-ray spectrum of AGNs modeled by a power-law component with both intrinsic and Galactic absorption (i.e., $zpow \times wabs \times zwabs$ in XSPEC) to estimate the intrinsic column density. A power-law photon index of $\Gamma=1.8$, which is typical for intrinsic AGN spectra, is assumed and the redshifts of the $zpow$ and $zwabs$ components are fixed to that of the source. We additionally fixed the Galactic column density to N$_{\rm H} = 6.0 \times 10^{19}$ cm$^{-2}$. We then derive the intrinsic column density that reproduces observed hard ($2-8$ keV) and soft ($0.5-2$ keV) band hardness ratios using XSPEC. The intrinsic X-ray luminosity is derived from the equation  L$_{\rm X}$ = 4$\pi$d$^{2}_{\rm L}~f_{\rm X, int}$ (1+z)$^{\Gamma -2}$ by correcting both intrinsic and Galactic absorption. $f_{\rm X, int}$ is the absorption-corrected X-ray flux and the d$_{\rm L}$ is luminosity distance. Finally, we derive the bolometric luminosity of AGNs from the absorption-corrected rest-frame intrinsic X-ray luminosity with the luminosity-dependent bolometric correction factor described in \citet{Marconi04}. They derived the bolometric corrections from an AGN template spectrum of optical-ultraviolet and X-ray luminosities radiated by the accretion disc and hot corona. They considered only the AGN accretion powered luminosity, neglecting the luminosity reprocessed by the dust, which is therefore representative of the AGN accretion power. The scatter is given by $\sim0.1$ for X-ray luminosities. 


\section{Black Hole Mass Estimation}

The black hole mass can be estimated using the broad-line width and the continuum (or line) luminosity from their single-epoch spectra as proxies for the characteristic velocity and the size of the broad-line region (e.g., \citealt{Kaspi00, Vestergaard02, Woo02, McLure02, McLure04, Greene05, Kollmeier06, Vestergaard06, Shen08, Shen11}). Depending on the redshift, single-epoch virial black hole masses have been estimated from different broad emission lines, such as \MgII~\citep{McLure02, McLure04, McGill08, Vestergaard09, Wang09, Shen11, Rafiee11}, \Hb~\citep{Greene05, Vestergaard06}, and \Ha~\citep{Greene05, Matsuoka13} lines. The virial black hole masses are calibrated against the black hole mass estimated by the reverberation mapping or that from the single-epoch broad-line width of \Hb~emission line in the local universe. Although there are several systematic uncertainties in these single-epoch virial black hole mass estimators, a number of studies have shown that there is consistency in black hole masses from various estimators. \citet{Shen12} point out that there is essentially no difference in black hole mass estimates using \MgII~and the Balmer lines for high redshift luminous AGNs. \citet{Matsuoka13} also show that virial black hole masses based on \Ha~and \MgII~emission lines are very similar over a wide range in black hole mass. They suggest that local scaling relations, using \Ha~or \MgII~emission lines, are applicable for moderate-luminosity AGNs up to $z\sim2$. 

We measure the properties of broad emission lines (e.g., \Ha, \Hb, and \MgII) present in optical and NIR spectra to derive single-epoch virial black hole mass of broad-line AGNs. The \Ha~$\lambda$6563\AA~and the \Hb~$\lambda$4861\AA~lines are redshifted to the NIR range, and the \MgII~$\lambda$2798\AA~line is present in optical spectra in the redshift range $0.5<z<2.5$. 

\subsection{Spectral line fitting}

We perform a fit to the emission lines using the mpfit routine, which adopts a Levenberg-Marquardt least-squares minimization algorithm to derive the best-fit parameters as well as a measure of the goodness of the overall fit. We specifically measure the width and the luminosity of emission lines in the case of \Ha~and \Hb~lines, and the width and the monochromatic continuum luminosity at 3000\AA~in the case of the \MgII~line. There might be a non-negligible host galaxy contribution at 3000\AA~continuum luminosity, but we do not correct for any contamination by the host galaxy and extinction due to dust. While we should be aware of this issue, the impact of these on black hole masses should be small, since the black hole mass scales with the square root of the luminosity (see Section 5.2).

Broad-line AGN spectra in the wavelength region of interest are usually characterized by a power-law continuum, $f_{\lambda} \propto \lambda^{-\alpha}$, and broad (and/or narrow) emission line components. We begin by fitting a power-law continuum with a slope of the continuum as a free parameter. In the case of the \MgII~line, it is crucial to consider a complex of \FeII~emission lines because in this wavelength range the lines are strongly blended with the broad \FeII~emission features (e.g. \citealt{Vestergaard01, Matsuoka07, Harris13}). We simultaneously fit the combination of a power-law continuum and \FeII~emission components. An empirical \FeII~emission template is adopted from \citet{Vestergaard01} and convolved with Gaussian profiles of various widths. We left the width, normalization, and offset from the line center as free parameters during the fit. From the best-fit power-law continuum, we derive an estimate of monochromatic luminosity at 3000\AA. Finally, we subtract the best-fit power-law continuum (and/or the \FeII~emission components) from the spectra. 

\begin{figure}
\centering
\includegraphics[width=0.5\textwidth]{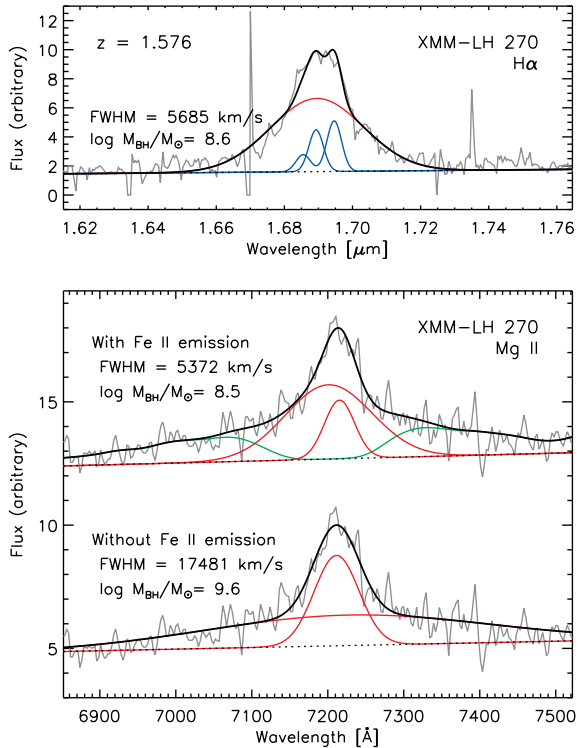}
\caption{Comparison of the broad-line fit for the \Ha~line (top panel) with that of the \MgII~line (bottom panel) with/without a \FeII~broad emission component for the same AGN source `{\it XMM}-LH 270' at $z=1.576$. The observed spectrum (grey) is shown with the best fit (black). In the top panel, the different components are shown as dotted lines (continuum), red curves (broad-line components), blue curves (narrow-line components of \Ha~and a pair of \NII~lines). In the bottom panel, the fit of the \MgII~line with \FeII~emission (upper) and without \FeII~emission (lower) are shown. The different components are indicated as red curves (individual broad-line components) and a green curve (\FeII~emission component).}
\end{figure}

\begin{figure}
\centering
\includegraphics[width=0.5\textwidth]{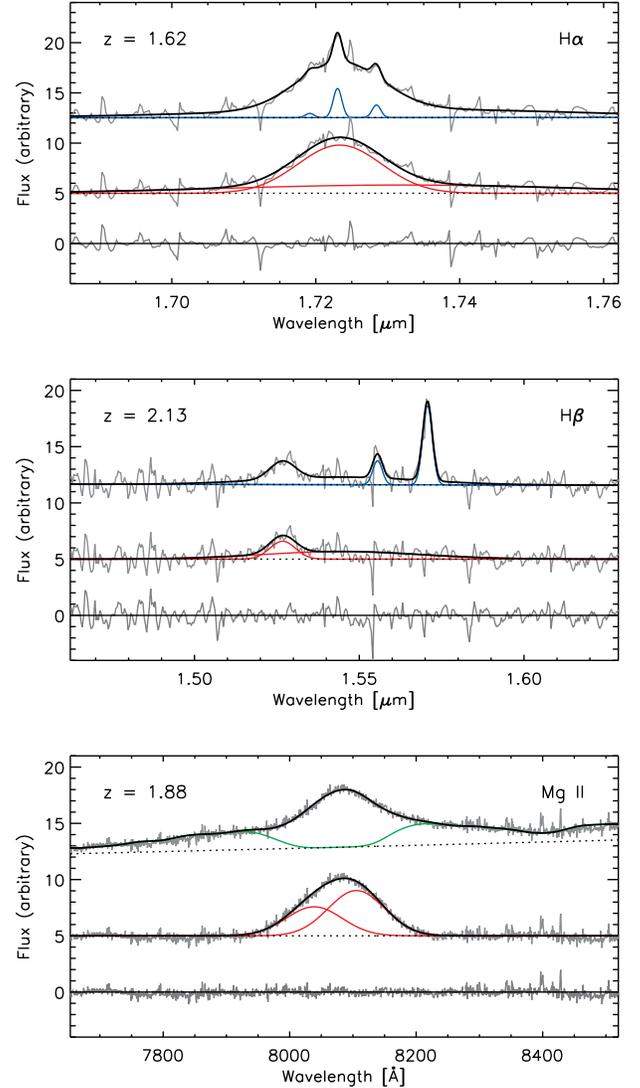}
\caption{Examples of the broad-line fits for \Ha~(top), \Hb~(middle), and \MgII~(bottom) emission lines at $z=1.62,~2.13,~{\rm and}~1.88$, respectively. The upper plot of each panel shows the observed spectrum (grey) with the best-fit model (black). The power-law continuum (dotted), narrow-line components (blue), and \FeII~emission component (green) are also indicated, respectively. The middle plot of each panel shows the only broad-line components after subtraction of the best-fit model of continuum, narrow components and \FeII~emission. The best-fit broad-line model is shown with the black curve. Each Gaussian broad-line component is also shown with red curves. The residual is shown in the lower plot of each panel.}
\end{figure}

We further consider individual components to determine the pure broad-line components that enable an accurate determination of the virial black hole masses. The line profile is described by a combination of multiple Gaussian components to best characterize the line shape in the sense that broad emission lines in AGNs can have a complex shape (e.g. Collin et al. 2006). The multiple Gaussian components provide non-Gaussian, asymmetric profiles reproducing the observed broad-line profile smoothly, but we are not concerned with the physical significance of the individual components. We fit the \Ha~$\lambda$6563\AA~(\Hb~$\lambda$4861\AA) line with one or two broad and a narrow Gaussian components, and the \NII~$\lambda$6548,6583\AA~(\OIII~$\lambda$4959,5007\AA) lines with a pair of Gaussians. The line ratio of the \NII~$\lambda$6548,6583\AA~and the \OIII~$\lambda$4959,5007\AA~lines are fixed to the laboratory value of 2.96 and 2.98, respectively. Both the narrow width of the \NII~and the \OIII~lines are fixed to match the narrow component of \Ha~and \Hb, respectively. We left the FWHM of the narrow line components as free parameters but limited to 900 km/s. For the \MgII~line, we fit with one or two broad Gaussians components. We do not consider the doublet component of the Mg II line because the line separation is small and does not affect the broad-line width. 

As a consistency check we compare the fit of the \MgII~line with \FeII~emission components to that of the \Ha~line, since the \Ha~line is not affected by \FeII~emission. In Figure 2, we show an example fit of the \Ha~line and that of the \MgII~line with/without \FeII~broad emission component for the same AGN source `{\it XMM}-LH 270' at $z=1.576$. We show the observed spectrum (grey) with the best fit (black) of the \Ha~line (top panel) and the \MgII~line (bottom panel). The different components are also indicated as red gaussian curves (broad-line components), blue curves (narrow-line components of \Ha~and a pair of \NII~lines), and green curve (\FeII~emission). While it is uncertain whether the \MgII~line is blended with \FeII~emission or it really has a very broad-component in the bottom panel, we confirm that the \MgII~line fit with \FeII~emission is likely to show a similar result with \Ha~line fit in the upper panel. 

In order to guarantee a reliable fit, we compare the fit with only narrow-line components, that with narrow-line and one broad Gaussian components, and that with narrow-line and two broad Gaussian components. We perform an F-test to decide whether an additional broad component is needed. We then subtract the narrow line components from the spectra obtaining a spectrum that contains only broad-line components. Finally, we inspect all fits by eye to check the cases where a broad component is unclear due to the low signal-to-noise ratio (S/N). We only consider spectra having S/N greater than 10 per pixel.

We determine the broad-line width and the line luminosity from the sum of the broad-line components. From the best-fit, the FWHM of the broad \Ha, \Hb, and \MgII~lines are computed and corrected for the effect of instrumental resolution to obtain an intrinsic velocity width. We select the broad-line AGNs with broad emission line widths larger than 2000 km/s of FWHM, a secure threshold for truly broadened lines, as compared to the spectral resolution. Additionally, we take into account the uncertainty in the derived FWHM and luminosity. We perform a Monte Carlo simulation comprising 100 realizations adding noise to each spectrum and iterate the whole procedure to find the best-fit model and the errors compatible with the observations, in order to assess the accuracy of the black hole mass measured. Since the best-fit model could have either one or two broad-line components during different Monte Carlo realizations for each spectrum, this could introduce a larger scatter.

In Figure 3, we show examples of broad-line fits for  \Ha~(top), \Hb~(middle), and \MgII~(bottom) emission lines at $z=1.62,~2.13,~{\rm and}~1.88$, respectively. The upper plot of each panel shows the observed spectrum (grey) with the best-fit model (black). The power-law continuum (black dotted), narrow-line components (blue), and \FeII~emission component (green) are also indicated, respectively. The middle plot of each panel shows the broad-line only components after subtraction of the best-fit model of continuum, narrow-line components and \FeII~emission. The best-fit broad-line model is shown with the black curve. Each Gaussian broad-line component is also shown with red curves. The residual is shown in the lower plot of each panel. 

\subsection{Black hole masses}

We calculate black hole masses from the FWHM and the luminosity of the sum of the broad line components. In the case of \Ha~and \Hb~we use the recipes provided by \citet{Greene05}. In addition, we specifically estimate the black hole mass based on the FWHM of the broad \MgII~line and the monochromatic continuum luminosity at 3000\AA~using the calibration derived by the \citet{McLure04}. The black hole mass can be expressed in the forms:

\begin{eqnarray}
\rm M_{BH} & = & 10^{6.301}(\frac{\rm L_{H\alpha}}{\rm 10^{42}~ergs~s^{-1}})^{0.55}(\frac{\rm FWHM_{H\alpha}}{\rm 10^{3}~km~s^{-1}})^{2.06}~\rm M_{\odot}
\end{eqnarray}
\begin{eqnarray}
\rm M_{BH} & = & 10^{6.556}(\frac{\rm L_{H\beta}}{\rm 10^{42}~ergs~s^{-1}})^{0.56}(\frac{\rm FWHM_{H\beta}}{\rm 10^{3}~\rm km~s^{-1}})^{2.0}~\rm M_{\odot}
\end{eqnarray}
\begin{eqnarray}
\rm M_{BH} & = & 10^{0.505}(\frac{\lambda \rm L_{\lambda3000}}{\rm 10^{44}~ergs~s^{-1}})^{0.62}(\frac{\rm FWHM_{MgII}}{\rm km~s^{-1}})^{2.0}~\rm M_{\odot} 
\end{eqnarray}

\noindent where FWHM is the FWHM of the line in units of 1000 km/s, and L$_{\lambda3000}$ is the continuum 
luminosity at 3000\AA. 

\begin{figure}
\centering
\includegraphics[width=0.5\textwidth]{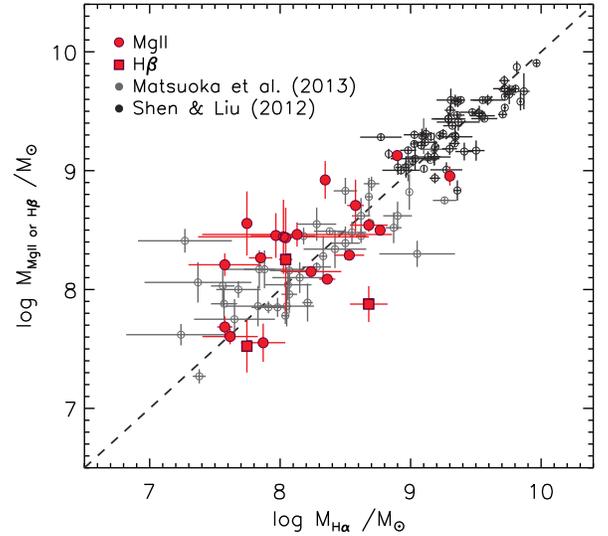}
\caption{Comparison of black hole masses estimated using the \Ha~line with that using the \MgII~line (red circles) or the \Hb~line (red squares). Our sample of AGNs are shown in red, and the observations from \citet{Matsuoka13} and \citet{Shen12} are also shown as grey and black symbols, respectively. The black dashed line denotes a one-to-one relation.}
\end{figure}

We present the comparison of black hole masses estimated using the \Ha~line with that using the \Hb~(red square), or \MgII~(red circles) lines in Figure 4, respectively. We also show the observations from \citet{Matsuoka13} and \citet{Shen12} as grey and black open circles for comparison, respectively. The black dashed line denotes a one-to-one relation. Our sample of broad-line AGNs spans a range of $7.0<{\rm log~M_{BH}/M_{\odot}}<9.5$ which is consistent with the previous studies of moderate-luminosity AGNs at $z\sim1-2$ \citep{Merloni10, Trump11, Matsuoka13}. The ratios of the mean black hole mass are log(M$_{\rm Mg\,\textsc{ii}}$)/ log(M$_{\rm H\alpha}$)=0.15, and log(M$_{\rm H\beta}$)/ log(M$_{\rm H\alpha})=-0.27$, respectively. The median uncertainty of the black hole mass is $\sim0.1$ dex. While there are offsets between the different black hole mass estimations, it is worth noting that the black hole mass estimated with different calibrations carries a scatter of $\sim0.3$ dex \citep{McGill08}. We also note that determination of black hole mass from the \Hb~emission line are known to be affected by significant systematic uncertainties due to the Balmer decrement. If there are multiple lines measured, we use the lines in order of \Ha, Mg II, and \Hb~for the determination of the black hole mass. There are six objects in our sample that black hole masses are determined with the \Hb~line.

\subsection{Broad-line AGNs}

We select the sample of broad-line AGNs for which one or more broad emission lines have been identified in the spectrum. From the NIR/optical spectra, the \Ha, \Hb,~and \MgII~wavelength regions are covered for 152, 56, and 62 spectra, respectively. Broad \Ha, \Hb,~and \MgII~lines are detected for 52, 7, and 53 in the NIR and/or optical spectra, respectively, by broad-line widths larger than 2000 km/s of FWHM with the high S/N. For 19 AGNs, broad lines are detected in both \Ha~and \MgII~lines (Figure 4). While all AGNs with detection of broad \Ha~lines are also detected in the broad \MgII~line, 5 AGNs with broad \MgII~line show no broad \Ha~line, mainly due to the low S/N NIR spectra. It is noted that there are quite a number clear broad \Ha~lines with practically absent \Hb~lines, indicating a large Balmer decrement. The final sample of broad-line AGNs in the CDF-S, E-CDF-S, and {\it XMM}-LH fields consists of 86 objects.

\begin{figure*}
\centering
\includegraphics[width=1\textwidth]{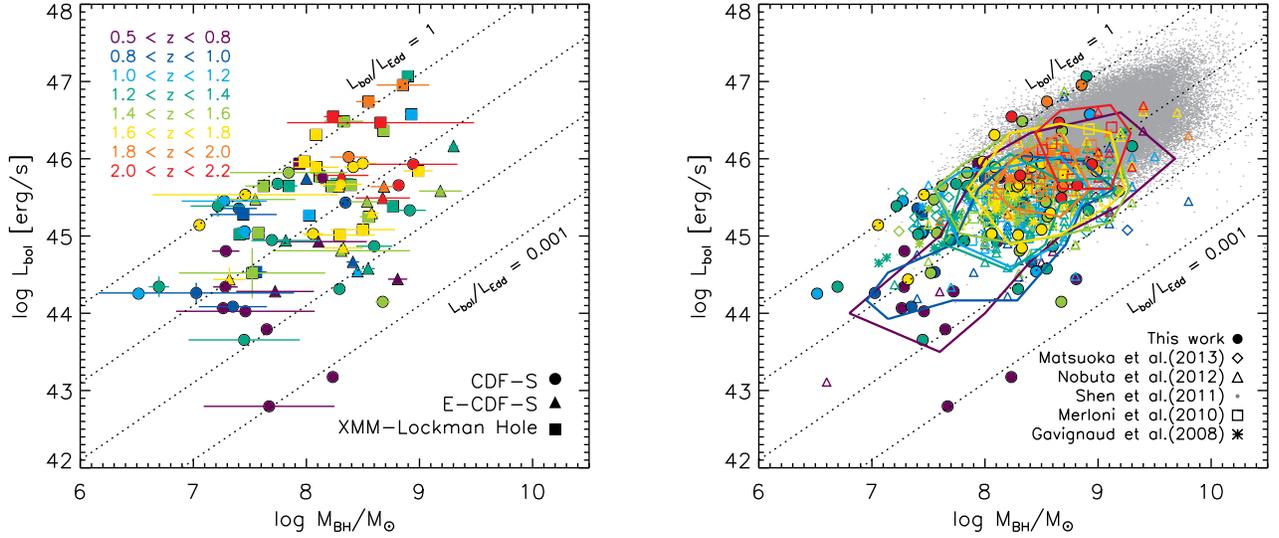}
\caption{AGN bolometric luminosity versus black hole mass for our sample of broad-line AGNs in the different redshift bins (left). In the right panel, contours at the 1$\sigma$ level are shown in the different redshift bins, together with the published observations from the literature as labeled. As a reference, lines of constant Eddington ratio (${\rm L_{bol}/L_{Edd}}$) equals to 1, 0.1, 0.01, and 0.001 are plotted as dotted lines.}
\end{figure*}


\section{Eddington Ratio distribution}

\begin{figure}
\centering
\includegraphics[width=0.5\textwidth]{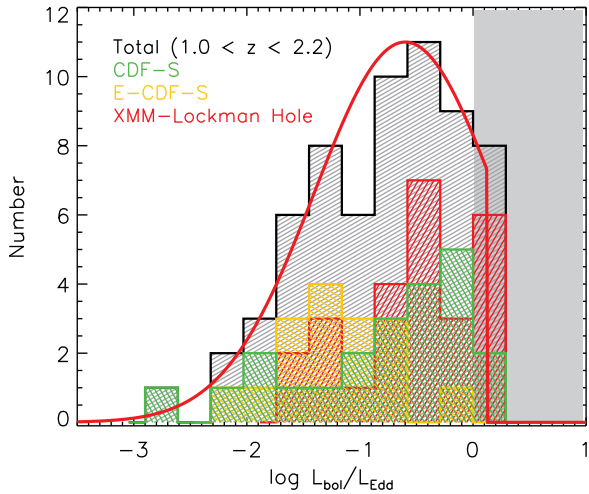}
\caption{Eddington ratio distribution of broad-line AGNs at $1.0<z<2.2$. The different X-ray surveys are shown in different color histograms, and the black histogram represents the combined distribution of all surveys. The grey shade indicates the Eddington limit. The red solid line indicates a log-normal fit with a peak of ${\rm log~L_{bol}/L_{Edd}=-0.6}$ and a dispersion of 0.8 dex.}
\end{figure}

The mass accretion onto the black hole is important for a better understanding of the AGN evolution. The Eddington ratio, the ratio between the AGN bolometric luminosity and the Eddington luminosity (L$_{\rm bol}$/L$_{\rm Edd}$), provides insight into the black hole growth since the bolometric luminosity reflects the mass accretion rate. We show AGN bolometric luminosity versus black hole mass for our sample of broad-line AGNs in the different redshift bins in the left panel of Figure 5. The different X-ray surveys are shown with different symbols as labeled. The dotted reference lines indicate constant Eddington ratios of 1, 0.1, 0.01, and 0.001, respectively. Our sample of broad-line AGNs covers the black hole mass range ${\rm 7.0<log~M_{BH}/M_{\odot}<9.5}$ and the bolometric luminosity range ${\rm 43<log~L_{bol}<47}$ with a wide dispersion in the Eddington ratio distribution. For comparison, we show published observations in the same redshift range from the literature in the right panel of Figure 5 \citep{Gavignaud08, Merloni10, Shen11, Nobuta12, Matsuoka13}. The SDSS quasar sample (grey points; \citealt{Shen11}) is limited to the high-mass and high-luminosity regime because the SDSS detection limit corresponds to a luminosity of ${\rm log~L_{bol}\sim46}$ at $z\sim1$. Compared to the SDSS quasar sample, our sample of broad-line AGNs show a wider dispersion in the black hole mass, AGN bolometric luminosity and Eddington ratio distribution, consistent with previous studies on deep AGN sample \citep{Gavignaud08, Merloni10, Nobuta12, Matsuoka13}, which fill in the low-mass and low-luminosity region. The figure shows contours at the 1$\sigma$ level, together with the literature data, except the SDSS quasar sample. The figure also reveals that only a small number of AGNs exceeds the Eddington limit by a small amount. AGNs with similar black hole masses show a broad range of bolometric luminosities spanning about two orders of magnitude, indicating that the accretion rate of black holes is widely distributed. This suggests that the AGN cosmic downsizing phenomenon could be explained by some more massive black holes with low accretion rates, which are relatively fainter than less massive black holes with efficient accretion. \citet{Lusso12} suggest that AGNs show higher Eddington ratios at higher redshift at any given ${\rm M_{BH}}$, and the Eddington ratio increases with bolometric luminosity. We confirm that there is a tendency for low-luminosity AGNs (${\rm log~L_{bol}\lesssim45.5}$) with less massive black holes (${\rm log~M_{BH}/M_{\odot}\lesssim8}$) to have lower Eddington ratios than high-luminosity AGNs (${\rm log~L_{bol}\gtrsim45.5}$) with massive black holes (${\rm log~M_{BH}/M_{\odot}\gtrsim8}$), consistent with \citet{Lusso12}. It is important to note that, when comparing with results in the literature, one should take into account the different methods of spectral line fitting and correction for bolometric luminosities. Nevertheless, they show similar distributions of the accretion rate of black holes over a wide range, consistent with previous studies. 

Several studies have found a correlation between the X-ray bolometric correction and the Eddington ratio (e.g. \citealt{Vasudevan07, Lusso12}), which may introduce biases into this diagram. \citet{Lusso12} found that there is a trend for higher bolometric corrections at higher bolometric luminosities. \citet{Vasudevan07} suggest that there appears to be a distinct step change in bolometric correction at an Eddington ratio of $\sim0.1$, below which apply lower bolometric corrections, and above which apply higher bolometric corrections. If one includes this correlation to the trend between bolometric luminosities and black hole masses in Figure 5, which low-luminosity AGNs have lower accretion rates while high-luminosity AGNs show higher accretion rates, would even be more pronounced. However, we note the possibility that there could be the spurious correlations since ${\rm L_{bol}}$ is present on both axes when plotting the bolometric correction against the Eddington ratio.

We show the Eddington ratio distribution of our sample of AGNs in the redshift range $1.0<z<2.2$ in Figure 6. The different X-ray surveys are shown in different color histograms, and the black histogram represents the combined distribution of all surveys. The distribution of Eddington ratios peaks at ${\rm log~L_{bol}/L_{Edd}\sim-1}$ with an extended tail towards low Eddington ratios, down to ${\rm log~L_{bol}/L_{Edd}\sim-3}$. A log-normal fit with a peak of ${\rm log~L_{bol}/L_{Edd}=-0.6}$ and a dispersion of 0.8 dex is shown as red solid line. In previous studies, \citet{Kollmeier06} suggest that the Eddington ratios are quite narrowly distributed independent of luminosity (${\rm L_{bol}=10^{45}-10^{47}~erg/s}$) and redshift ($0.3<z<4.0$), with a dispersion of 0.3 dex (see also \citealt{Steinhardt10}). \citet{Lusso12} also suggest that the distribution of Eddington ratios are nearly Gaussian especially at high-redshift and at high ${\rm L_{bol}/M_{BH}}$, with a dispersion of $\sim$0.35 dex, while the low-redshift and low ${\rm L_{bol}/M_{BH}}$ are more affected by incompleteness. We list our sample of broad-line AGNs in Appendix Table 3, which includes AGN bolometric luminosities, black hole masses, and measurements of emission line properties.

We should emphasize here that the systematic selection effects could certainly be playing a role in determining the distribution of AGN bolometric luminosities and black hole masses. The Eddington ratio distribution, thus, could be a result of the selection bias, mainly the limited X-ray luminosity but also to the broad line width, i.e. the black hole mass. The X-ray luminosity is limited by the X-ray flux limit, depending on redshift and on the limited volume. The detectability of the broad emission line gives rise to a bias against the black hole mass. Also, the black hole mass could be biased by observational limitations to detect the corresponding very broad lines and low signal-to-noise spectra. This is bound to introduce selection biases, which could mimic artificial correlations in the data. Hence, we will further discuss the possible selection effects in the next section.

\begin{figure*}
\centering
\includegraphics[width=1\textwidth]{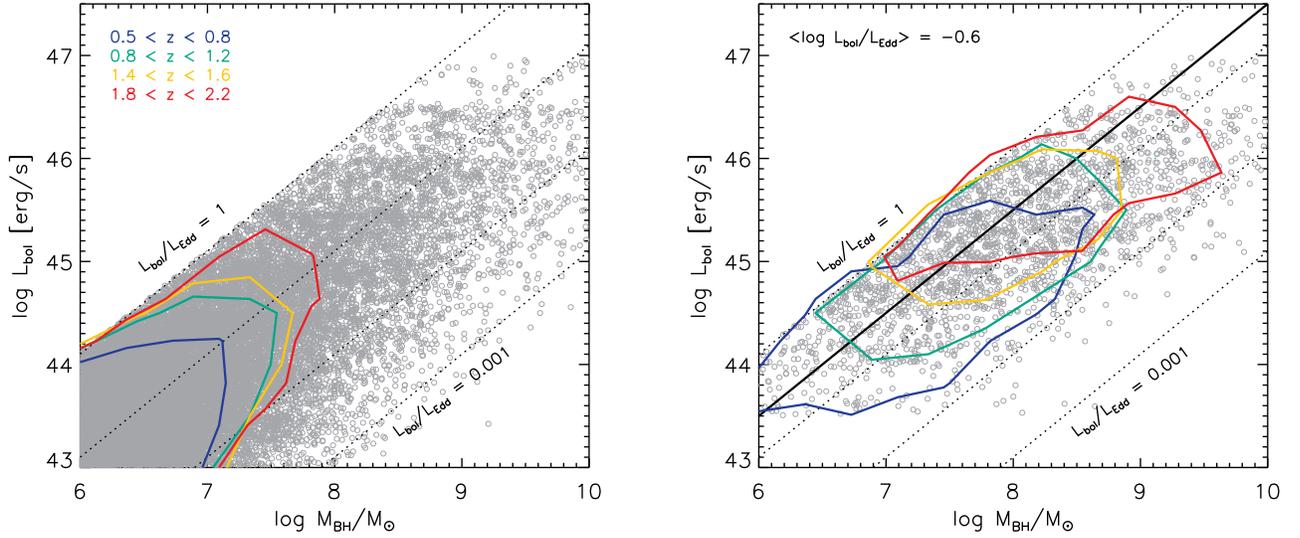}
\caption{Monte Carlo simulated data sets from the AGN bolometric luminosity function \citep{Hopkins07} in the different redshift bins with an assumption for the Eddington ratio distribution, which has a peak of ${\rm log~L_{bol}/L_{Edd}=-0.6}$ and a dispersion of 0.8 dex (red curve in Figure 6) regardless of AGN luminosity or redshift. The simulated data sets (left), and those which are affected by the same observed selection effects (right) are shown in grey. As a reference, lines of constant Eddington ratio (${\rm L_{bol}/L_{Edd}}$) equals to 1, 0.1, 0.01, 0.001 are plotted as dotted lines. Contours at the $1\sigma$ level are shown in the different redshift bins. The black solid line indicates the assumed peak of Eddington ratio.}
\end{figure*}

\begin{figure*}
\centering
\includegraphics[width=1\textwidth]{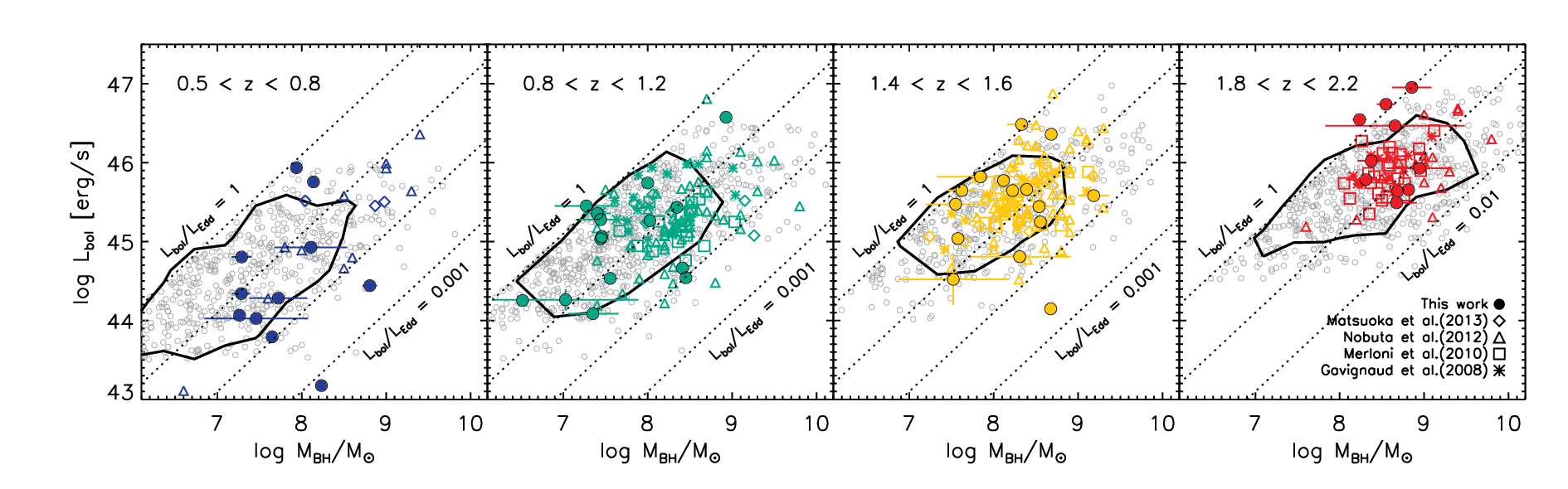}
\includegraphics[width=1\textwidth]{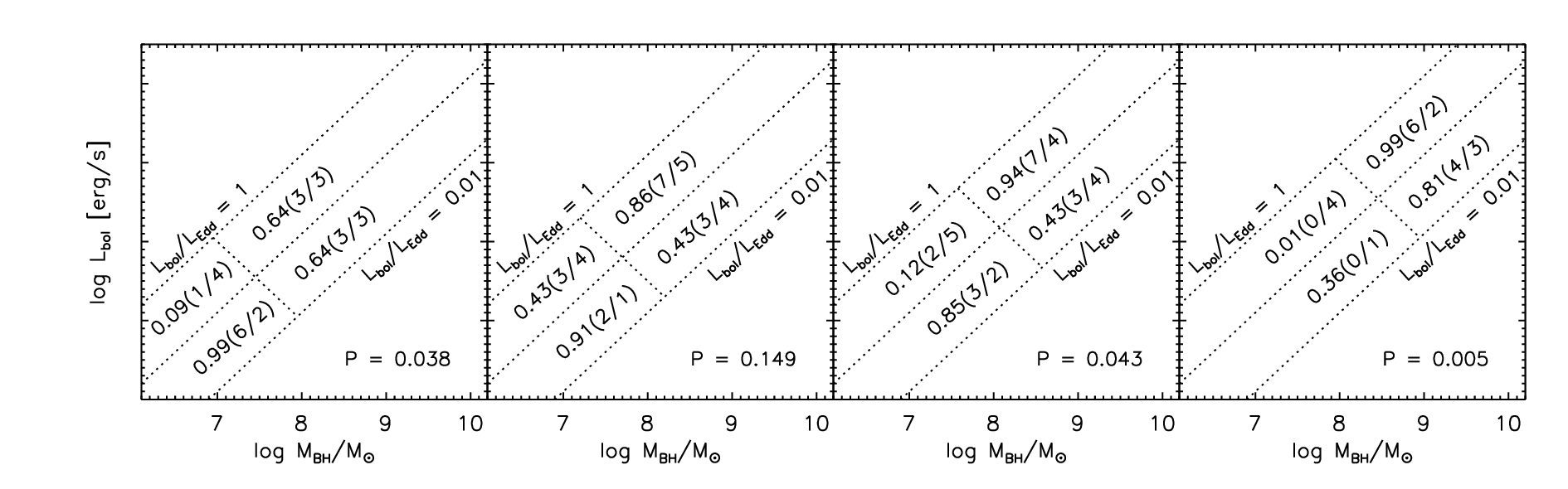}
\caption{Comparison of the Monte Carlo simulated data sets with the observed AGNs in the different redshift bins. In the top panels, the simulated data sets are shown in grey with the $1\sigma$ contour, and the observed AGNs are shown as purple ($0.5<z<0.8$), green ($0.8<z<1.2$), yellow ($1.4<z<1.6$), and red ($1.8<z<2.2$) symbols, respectively. In the bottom panels, the Poisson likelihood is shown in each bin, which is perpendicular to the Eddington ratio plane. The numbers in parentheses refer to detected (observed) sources over expected (simulated) sources. Reference lines of constant Eddington ratios are plotted as dotted lines.}
\end{figure*}


\section{Analysis of selection biases}

We investigate the possible bias due to systematics and selection effects on the observed AGN bolometric luminosity and the black hole mass. To explore the effect of these selection biases, we construct Monte Carlo simulations to make artificial datasets, which are affected by the same selection effects. We start from the bolometric luminosity function of AGNs \citep{Hopkins07} in the different redshift bins with an assumption for the Eddington ratio distribution, which has a peak of ${\rm log~L_{bol}/L_{Edd}=-0.6}$ and a dispersion of 0.8 dex, bounded by ${\rm L_{Edd}}$, taken from the observed distribution (see red curve in the top panel of Figure 6). To account for the observed selection biases, we apply the same selection effects based on our combined X-ray surveys. The X-ray flux limit corresponds to a bolometric luminosity of ${\rm log~L_{bol}\sim43}$ at $z\sim1$. We use the survey area of the total combined X-ray surveys (black curve in the right panel of Figure 1). Since it has been known that there is much larger fraction of obscured AGN at lower luminosities \citep{Ueda03, Steffen03, Simpson05, Hasinger08}, we apply the fraction of broad-line AGNs as a function of AGN luminosity from \citet{Hasinger08}, to which the same bolometric correction (\citealt{Marconi04}) was applied to the X-ray luminosity. The black hole mass is biased by the detectability of the broad emission line and the signal-to-noise of spectra. We, thus, apply a `fudge' factor, which is the exponentially decaying function at low-mass (${\rm 6.5<log~M_{BH}<7.5}$) and high-mass (${\rm 8.5<log~M_{BH}<9.5}$). The `fudge' factor brings down the numbers of low-mass (${\rm 6.5<log~M_{BH}<7.5}$) and high-mass (${\rm 8.5<log~M_{BH}<9.5}$) AGNs, which takes into account the spectroscopic incompleteness. For each data set of the different redshift bins, we calculate black hole masses from the AGN bolometric luminosity and the Eddington ratio. The Eddington ratio distribution is assumed to be same regardless of AGN luminosity or redshift, which is a valid assumption for the high-luminosity AGNs \citep{Kollmeier06}. \citet{Steinhardt10} report that the Eddington ratio distributions are all similar for SDSS quasar populations over a wide range of mass and redshift.

In Figure 7, the Monte Carlo simulated data sets (left panel), and those which are affected by the same observed selection effects (right panel) are shown with grey symbols and contours at the $1\sigma$ level in the different redshift bins. The black solid line in the right panel of Figure 7 indicates the assumed peak of Eddington ratio, ${\rm log~L_{bol}/L_{Edd}=-0.6}$. The AGN downsizing trend is seen in the sense that the characteristic AGN luminosity and black hole mass decrease with redshift. This is primarily due to the strong evolution of the co-moving number density at the bright end of the AGN luminosity function at $0.5<z<2.0$, together with the corresponding selection biases. We compare the simulated data with the observed AGNs in the different redshift bins in Figure 8. The simulated data sets are shown in grey with the contour at the $1\sigma$ level, and the observed AGNs are shown in colored symbols for each redshift bin in the top panels. For each redshift bin we group the data into four sets, using a constant Eddington ratio of ${\rm L_{bol}/L_{Edd}=0.1}$ and a line perpendicular to the Eddington ratio as separation. In the bottom panels, the number of detected (observed) sources over the number of expected (simulated) sources is given in parentheses, as well as the Poisson likelihood calculated from this combination. In bins of high-luminosity AGNs with high Eddington ratio as well as low-luminosity AGNs with low Eddington ratio the detected number of objects agrees with the prediction from the Monte Carlo simulation within the statistical errors. However, for low-luminosity AGNs with high Eddington ratios, especially at high redshift ($1.8<z<2.2$) and low redshift ($0.5<z<0.8$) bins, the simulations systematically predict a larger number of objects, than those observed. Taking all Poisson likelihoods together, there is a difference between the observed and the predicted distributions. We therefore suggest that there is a dependence of AGN luminosities on the Eddington ratios in the sense that luminous AGNs appear to have systematically higher Eddington ratios than low-luminosity AGNs. However, this result is of marginal significance due to the relatively small number of objects in each bin. We note that our sample of high-luminosity X-ray selected AGNs overlaps with the less luminous quasars from the SDSS sample at the highest Eddington ratios (see grey points in the right panel of Figure 5), while the most luminous, most massive SDSS quasars lie further away from their Eddington luminosity \citep{Steinhardt10}.


\section{Discussion}

We now discuss the observed AGN downsizing phenomenon, and possible explanations for the black hole growth over cosmic time. The decrease of the characteristic luminosity of AGNs with redshift has been described as AGN downsizing, implying that the AGN activity at earlier epochs was much more intense. We show that AGNs with similar black hole masses show a broad range of bolometric luminosities, which means the accretion rate of black holes is widely distributed. The average accretion rate of two different AGN fueling mechanisms can play a crucial role for the downsizing interpretation. AGN activity triggered by major mergers is thought to have a higher accretion rate than activity triggered by secular evolution effects. Therefore the luminosity of an AGN with a certain black hole mass may differ widely, depending on the accretion mechanisms. 

The colors and morphologies of galaxies may contain a record of their growth history. Several studies have addressed that the majority of AGN host galaxies in the local universe are preferentially in the ``green valley" on the color-magnitude diagram, between actively star-forming galaxies in the blue cloud and passively evolving galaxies on the red sequence (e.g. \citealt{Schawinski10}). Moreover, a large fraction of moderate-luminosity AGNs are likely to live in disk-dominated galaxies \citep{Gabor09, Cisternas11, Schawinski11, Kocevski12, Mullaney12, Rosario15}. \citet{Fan14} find that the majority of AGN host galaxies show no significant merger features up to $z\sim2$. It is likely that merger features are visible only for a few Gyrs after major mergers (e.g. \citealt{Lotz08, Ji14}), suggesting that most AGN activity does not seem to be triggered by major mergers since $z\sim2$ (see also \citealt{Mainieri11, Silverman11, Schawinski12, Schramm13, Villforth14}). \citet{Allevato11} further point out that moderate-luminosity AGNs at $z=0-2$ live in relatively massive dark matter halos (${\rm 10^{13.5}M_{\odot}}$), which corresponds to rich groups of galaxies, independent of redshift. The rich group environment may provide a kind of ``goldilocks" zone for AGNs in the sense that on one hand the density is high enough to cause frequent gravitational disturbances bringing cold gas to the center, and on the other hand the gas density in the group is not high enough to remove the cold gas from the galaxies due to ram pressure stripping. This also indicates that major mergers cannot be the main driver of the late evolution of AGNs. This raises interesting questions regarding different fueling mechanisms for the growth of black holes and galaxies at different epochs during cosmic time.

Given these intriguing findings, a possible interpretation for explaining the cosmic downsizing as well as morphologies and colors of AGN host galaxies is that there are two different modes of AGN feedback at different epochs (see \citealt{Hasinger08}). In an active AGN phase at high redshift, black holes have experienced vigorous growth by major mergers while radiating close to the Eddington limit (see e.g. \citealt{DiMatteo05}). When they reach a critical mass, at which the AGN is sufficient to blow out the surrounding gas, the feedback of the black hole suppresses further star formation and creates a red bulge-dominated remnant (e.g. \citealt{Fabian99, Springel05}). It is likely that only a small fraction of the transient population can be found in between the blue cloud and the red sequence due to rather short merger timescale ($\sim10^{8}$ yrs). The relatively massive galaxies, which have already experienced substantial growth by previous mergers, grow slowly through episodic star formation via secular evolution, leading to a disc surrounding the bulge. The modest AGN activity can be triggered by the gas accretion over cosmic time via internal, secular processes, such as gravitational instabilities in the disc. This secular growth is slow enough, and thus, the presence of AGN host galaxies in the green valley on the color-magnitude diagram could be interpreted as evidence for the on-going star formation in the inner region of low-luminosity AGN host galaxies at lower redshift, coming down from the red sequence. This is also compatible to the weak link between merger features and the AGN activity, as well as the moderate-luminosity AGNs in the relatively massive dark matter halos at $z\lesssim2$, where the number density of most luminous AGNs starts to decline. Finally, the late feedback from AGNs suppresses the late cooling flows of hot gas, keeping the galaxy quiescent. This seems to consistent with dormant supermassive black holes in dynamically hot systems (e.g. massive early-type galaxies) that contain little cold gas and correspondingly little star-formation. All of these seem to be consistent with the hierarchical growth scenario.


\section{Summary}

We present the Eddington ratio distribution of X-ray selected broad-line AGNs in the CDF-S, E-CDF-S, and the {\it XMM}-LH surveys. We calculate AGN bolometric luminosities from absorption-corrected X-ray luminosities, and estimate black hole masses of broad-line AGNs using the optical and Subaru/FMOS near-infrared spectroscopy. Our sample of broad-line AGNs spans the bolometric luminosity range ${\rm L_{bol}\sim10^{43.5-47}~erg/s}$, and the black hole mass range ${\rm M_{BH}\sim10^{6.5-9.5}~M_{\odot}}$ with a broad range of Eddington ratios ${\rm L_{bol}/L_{Edd}\sim0.001-1}$.

We explore the systematics and selection biases, because in general observed distributions are dependent on the X-ray flux limit and the detectability of the broad emission lines. Based on the analysis on these effects, we find that the observed downsizing trend could be simply explained by the strong evolution of the co-moving number density at the bright end of the AGN luminosity function at $0.5<z<2.0$, together with the corresponding selection effects. However, in order to explain the relatively small fraction of low-luminosity AGNs with high accretion rates, we might need to consider a correlation between the AGN luminosity and the accretion rate of black holes that luminous AGNs have higher Eddington ratios than low-luminosity AGNs. We suggest that the AGN downsizing trend can be interpreted as the fraction of AGNs radiating close to the Eddington limit decrease after their peak activity phases, suggesting that the fueling mechanism of growth of black holes might change through the cosmic time.


\acknowledgments

We thank the anonymous referee for several comments, which helped to improve the quality of the manuscript significantly. 
We thank Amy Barger for useful comments that helped improve this paper.

{\it Facilities:} \facility{Subaru (FMOS)}


\appendix

\section{Subaru/FMOS spectroscopic observations}

In Table 2, we present the identified spectroscopic redshifts from the Subaru/FMOS observations. The ID is from the published catalog of CDF-S \citep{Xue11}, E-CDF-S \citep{Lehmer05}, and {\it XMM}-LH \citep{Brunner08}. We assign a quality flag that gives the confidence in the redshift measurement. Flag 2 indicates a reliable redshift due to high signal-to-noise ratio (S/N) spectra and multiple spectral features. Flag 1 indicates that a redshift is not securely identified due to either low S/N or the presence of only a single emission line with no additional features. The newly discovered spectroscopic redshifts are marked with a cross sign. ``++" if no previous spectroscopic redshift and ``+" if the previous spectroscopic redshift is insecure. The redshift column gives our best redshift estimate. We also present the previous spectroscopic and photometric redshifts from the catalog.

In Table 3, we list our sample of broad-line AGNs, which includes AGN bolometric luminosities, black hole masses, and measurements of emission line properties.

\begin{longtable*}{l r r r c l c c l l}
  \caption{Subaru/FMOS spectroscopic observations} \\
 \hline \hline
     \multicolumn{1}{c}{\textbf{Field}} & 
    \multicolumn{1}{c}{\textbf{ID}} &
    \multicolumn{1}{c}{\textbf{RA}} &
    \multicolumn{1}{c}{\textbf{Dec}} & 
    \multicolumn{1}{c}{\textbf{Redshift}} &
    \multicolumn{1}{c}{\textbf{Quality}} &
    \multicolumn{1}{c}{\textbf{z$_{\rm \bf spec}$}} &
    \multicolumn{1}{c}{\textbf{z$_{\rm \bf phot}$}} &
    \multicolumn{1}{c}{\textbf{Band}} &
    \multicolumn{1}{c}{\textbf{Emission features}}\\ \hline \\[-1.8ex]
\endfirsthead
\multicolumn{10}{c}{\tablename \thetable{} -- continued.} \\[0.5ex]
\hline \hline \\[-1.8ex]
    \multicolumn{1}{c}{\textbf{Field}} & 
    \multicolumn{1}{c}{\textbf{ID}} & 
    \multicolumn{1}{c}{\textbf{RA}} &
    \multicolumn{1}{c}{\textbf{Dec}} & 
    \multicolumn{1}{c}{\textbf{Redshift}} &
    \multicolumn{1}{c}{\textbf{Quality}} &
    \multicolumn{1}{c}{\textbf{z$_{\rm \bf spec}$}} &
    \multicolumn{1}{c}{\textbf{z$_{\rm \bf phot}$}} &
    \multicolumn{1}{c}{\textbf{Band}} &
    \multicolumn{1}{c}{\textbf{Emission features}}\\ \hline \\[-1.8ex]
\endhead
\\[-1.8ex] \hline \hline
\endfoot
  CDF-S &    1 &  52.899 & -27.860 & 1.630 & 2+  & 1.624 & 1.626 & J-long, H-long & \Hb, \Ha, \NII, \SII \\
  CDF-S &    4 &  52.930 & -27.901 & 1.270 & 2+  & 1.271 & 1.027 & J-long, H-short, H-long & \Ha, \SII \\
  CDF-S &     7 & 52.936 & -27.865 & 0.880 & 1++ & -99.0 & 0.881 & H-long & \SII \\
  CDF-S &   25 &  52.960 & -27.870 & 1.336 & 2 & 1.374 & 1.386 & H-short & \Ha, \SII \\
  CDF-S &   26 &  52.960 & -27.864 & 1.447 & 1++ & -99.0 & 1.438 & J-long, H-short, H-long & \Hb \\
  CDF-S &   31 &  52.963 & -27.744 & 1.608 & 2++ & -99.0 & 1.829 & H-short, H-long & \Ha, \NII, \SII \\
  CDF-S &   36 & 52.967 & -27.804 & 1.236 & 1+ & 1.236 & 1.223 & H-short & \Ha, \NII \\  
  CDF-S &   37 &  52.968 & -27.696 & 1.020 & 2+ & 0.857 & 1.021 & J-long & \Ha, \NII \\
  CDF-S &   49 &  52.980 & -27.841 & 1.212 & 1+ & 1.212 & 1.208 & H-short & \NII, \SII \\
  CDF-S &   51 &  52.981 & -27.913 & 0.726 & 2 & 0.737 & 0.724 & J-long & \Ha, \SII \\
  CDF-S &   54 &  52.983 & -27.823 & 1.219 & 2++ & -99.0 & 1.208 & H-short & \Ha, \NII \\
  CDF-S &   64 &  52.991 & -27.793 & 0.856 & 1++ & -99.0 & -99.0 & J-long, H-long & \Ha \\
  CDF-S &   67 &  52.993 & -27.845 & 1.544 & 2++ & -99.0 & 3.309 & J-long, H-long & \Ha \\
  CDF-S &   72 & 52.998 & -27.839 & 1.638 & 1+ & 1.574 & 1.900 & J-long, H-long & \Ha \\  
  CDF-S &   76 &  53.002 & -27.722 & 1.042 & 2 & 1.037 & 1.05 & J-long, H-long & \Ha, \NII \\
  CDF-S &   80 &  53.003 & -27.893 & 1.549 & 2++ & -99.0 & 1.593 & J-long, H-short, H-long & \Ha, \NII \\
  CDF-S &   81 &  53.004 & -27.799 & 0.975 & 2 & 0.975 & 0.951 & J-long & \Ha, \NII, \SII \\
  CDF-S &   83 &  53.006 & -27.780 & 1.307 & 2++ & -99.0 & 1.499 & H-short, H-long & \Ha \\
  CDF-S &   84 &  53.006 & -27.694 & 1.414 & 1 & 1.406 & 1.379 & H-short & \Ha, \NII \\
  CDF-S &   88 &  53.010 & -27.767 & 1.613 & 1 & 1.616 & 1.626 & J-long, H-long & \Hb, \Ha \\
  CDF-S &   99 &  53.016 & -27.891 & 3.039 & 1++ & -99.0 & 3.877 & H-short & \OII \\
  CDF-S &  101 &  53.017 & -27.624 & 0.966 & 2 & 0.977 & 1.000 & J-long & \Ha \\
  CDF-S &  109 &  53.020 & -27.691 & 1.524 & 2+ & 0.720 & 1.562 & J-long, H-long & \Hb, \OIII \\
  CDF-S &  112 &  53.023 & -27.757 & 1.189 & 1+ &1.189 & 1.129 & H-short & \Ha, \NII \\
  CDF-S &  113 &  53.024 & -27.746 & 1.609 & 1+ & 1.608 & 1.692 & J-long, H-long & \OII, \Ha \\
  CDF-S &  117 & 53.025 & -27.824 & 1.295 & 1++ & -99.0 & 1.208 & J-long, H-short, H-long & \Ha \\
  CDF-S &  118 & 53.026 & -27.673 & 1.624 & 2++ & -99.0 & 1.438 & H-long & \Ha, \NII \\
  CDF-S &  120 &  53.027 & -27.791 & 1.021 & 2 & 1.021 & 1.023 & J-long & \Ha, \NII \\
  CDF-S &  121 &  53.027 & -27.765 & 1.329 & 1 & 1.329 & 1.321 & H-short & \Ha, \NII \\
  CDF-S &  127 &  53.029 & -27.936 & 0.774 & 2 & 0.777 & 0.767 & J-long, H-long & \Ha, \NII, \SII, \SIII \\
  CDF-S &  131 &  53.031 & -27.856 & 1.546 & 1+ & 0.034 & 1.546 & J-long, H-long & \OIII, \Ha \\
  CDF-S &  135 &  53.033 & -27.626 & 0.977 & 1 & 0.976 & 1.000 & J-long, H-short, H-long & \Ha, \NII \\
  CDF-S &  143 &  53.036 & -27.850 & 0.736 & 2 & 0.736 & 0.718 & J-long & \Ha, \NII \\
  CDF-S &  152 &  53.041 & -27.887 & 0.740 & 1 & 0.743 & 0.750 & J-long & \Ha, \SII \\
  CDF-S &  165 &  53.046 & -27.729 & 0.998 & 2 & 0.998 & 1.000 & J-long & \Ha, \NII \\
  CDF-S &  166 &  53.046 & -27.738 & 1.611 & 1 & 1.605 & 1.626 & J-long, H-long & \Ha, \NII \\
  CDF-S &  182 &  53.052 & -27.820 & 1.048 & 2+ & 1.048 & 0.904 & J-long & \Ha \\
  CDF-S &  205 &  53.060 & -27.853 & 1.540 & 2 & 1.544 & 1.562 & J-long, H-long & \OIII, \Ha, \NII \\
  CDF-S &  222 &  53.066 & -27.702 & 1.691 & 1++ & -99.0 & 2.779 & H-long & \Ha \\
  CDF-S &  226 &  53.067 & -27.817 & 1.412 & 1 & 1.413 & 1.438 & H-short & \Ha \\
  CDF-S &  229 &  53.068 & -27.658 & 1.326 & 2 & 1.324 & 1.321 & J-long, H-short & \Hb, \OIII, \Ha, \NII, \SII \\
  CDF-S &  236 &  53.071 & -27.834 & 1.613 & 2 & 1.611 & 1.579 & J-long, H-long & \Ha, \SII \\
  CDF-S &  257 &  53.076 & -27.849 & 1.535 & 2 & 1.536 & 1.562 & H-long & \Ha \\
  CDF-S &  271 &  53.081 & -27.681 & 0.681 & 1 & 0.761 & 0.709 & H-short & \SIII \\
  CDF-S &  290 &  53.087 & -27.930 & 2.301 & 2++ & -99.0 & 2.826 & H-short & \Hg \\
  CDF-S &  308 &  53.094 & -27.768 & 1.729 & 1 & 1.735 & 1.264 & J-long, H-long & \OIII, \Ha \\
  CDF-S &  329 &  53.102 & -27.670 & 0.954 & 2++ & -99.0 & 1.050 & J-long & \Ha, \NII \\
  CDF-S &  344 &  53.105 & -27.705 & 1.615 & 2 & 1.617 & 1.626 & J-long, H-long & \OIII, \Ha, \NII, \SII \\
  CDF-S &  349 &  53.106 & -27.772 & 0.896 & 2 & 0.896 & 0.904 & J-long & \Ha, \NII \\
  CDF-S &  368 &  53.111 & -27.824 & 1.469 & 2 & 1.468 & 1.560 & J-long, H-long & \OIII, \Ha, \SII \\
  CDF-S &  369 &  53.111 & -27.670 & 1.612 & 2 & 2.208 & 2.202 & J-long, H-long & \Hb, \Ha, \NII, \SII \\
  CDF-S &  405 &  53.123 & -27.723 & 1.614 & 2+ & 1.609 & 1.598 & J-long, H-short, H-long & \Ha, \NII \\
  CDF-S &  409 &  53.124 & -27.863 & 1.316 & 2+ & 1.316 & 5.247 & H-short & \Ha \\
  CDF-S &  414 &  53.124 & -27.756 & 0.953 & 2 & 0.953 & 0.951 & J-long & \Ha \\
  CDF-S &  415 &  53.125 & -27.717 & 1.357 & 2 & 1.356 & 1.328 & H-short & \Ha, \NII, \SII \\
  CDF-S &  417 &  53.125 & -27.758 & 1.222 & 2 & 1.209 & 1.208 & H-short & \Ha, \NII, \SII \\
  CDF-S &  420 &  53.125 & -27.756 & 0.960 & 2 & 0.960 & 0.951 & J-long & \Ha, \NII, \SII \\
  CDF-S &  424 &  53.126 & -28.005 & 2.310 & 2 & 2.306 & 2.282 & J-long, H-long & \OIII \\
  CDF-S &  429 &  53.130 & -27.655 & 1.040 & 1+ & 1.038 & 1.438 & J-long, H-long & \Ha, \NII \\
  CDF-S &  436 &  53.131 & -27.841 & 1.549 & 1+ & 1.613 & 1.485 & J-long, H-long & \Ha, \NII \\
  CDF-S &  439 &  53.132 & -27.833 & 0.980 & 1 & 0.98 & 0.983 & J-long & \Ha \\
  CDF-S &  467 &  53.142 & -27.841 & 1.384 & 2 & 1.384 & 1.395 & H-short & \Ha \\
  CDF-S &  473 &  53.144 & -27.654 & 1.557 & 2 & 1.510 & 1.499 & H-long & \Ha, \NII, \SII \\
  CDF-S &  487 & 53.147 & -27.667 & 1.473 & 1++ & -99.0 & 1.490 & J-long, H-short, H-long & \Ha \\ 
  CDF-S &  491 & 53.149 & -27.792 & 1.223 & 1 & 1.223 & 1.208 & H-short & \SII \\  
  CDF-S &  492 &  53.149 & -27.683 & 0.735 & 2 & 0.735 & 0.724 & J-long & \Ha, \NII, \SII \\
  CDF-S &  499 &  53.151 & -27.857 & 1.614 & 2 & 1.613 & 1.626 & H-long & \Ha \\
  CDF-S &  503 & 53.151 & -27.713 & 1.542 & 2+ & 1.609 & 1.582 & J-long, H-long & \Ha, \NII, \SII \\ 
  CDF-S &  518 &  53.157 & -27.870 & 1.611 & 2 & 1.603 & 1.692 & J-long, H-long & \OIII, \Ha, \NII, \SII \\
  CDF-S &  519 &  53.158 & -27.704 & 1.791 & 1+ & 1.814 & -99.0 & J-long, H-long & \Hg \\
  CDF-S &  524 &  53.160 & -27.931 & 2.043 & 1++ & -99.0 & 1.829 & J-long, H-short, H-long & \OIII \\
  CDF-S &  557 & 53.171 & -27.741 & 1.299 & 1 & 1.298 & 1.142 & H-short & \Ha, \NII \\ 
  CDF-S &  580 & 53.181 & -27.783 & 1.571 & 1+ & 1.570 & 1.578 & J-long, H-long & \Hb, \Ha, \NII \\   
  CDF-S &  586 &  53.184 & -27.793 & 0.738 & 2 & 0.738 & 0.724 & J-long & \Ha \\
  CDF-S & 599 & 53.188 & -27.904 & 1.380 & 1 & 1.378 & 1.328 & H-short & \Ha, \NII \\  
  CDF-S & 612 & 53.192 & -27.891 & 1.381 & 1+ & 1.382 & 1.428 & H-short & \Ha, \NII, \SII \\
  CDF-S &  619 &  53.196 & -27.730 & 1.212 & 1 & 1.178 & 1.154 & H-short & \Ha, \NII \\
  CDF-S &  623 &  53.197 & -27.713 & 0.732 & 1 & 0.729 & -99.0 & J-long, H-long & \Ha, \SII \\
  CDF-S &  626 &  53.200 & -27.709 & 0.979 & 2 & 0.979 & 1.000 & J-long, H-long & \Ha, \NII, \SII \\
  CDF-S &  629 &  53.201 & -27.882 & 0.660 & 1 & 0.667 & 0.601 & H-short & \SIII \\
  CDF-S &  634 &  53.205 & -27.681 & 1.226 & 2 & 1.222 & 1.208 & J-long, H-short & \Ha, \NII \\
  CDF-S &  648 & 53.214 & -27.929 & 1.448 & 2++ & 0.853 & 0.857 & J-long, H-short, H-long & \SII \\ 
  CDF-S &  652 &  53.216 & -27.708 & 1.023 & 2 & 1.020 & 1.000 & J-long & \Ha, \NII \\
  CDF-S &  656 &  53.218 & -27.762 & 1.367 & 2 & 1.367 & 1.379 & H-short & \Ha, \NII, \SII \\
  CDF-S &  661 &  53.226 & -27.818 & 0.987 & 1+ & 0.988 & 0.999 & J-long & \Ha \\
  CDF-S &  680 &  53.246 & -27.861 & 1.502 & 2++ & -99.0 & 1.379 & J-long, H-short, H-long & \Ha, \NII, \SII \\
  CDF-S &  683 &  53.247 & -27.816 & 1.613 & 2++ & -99.0 & 2.202 & J-long, H-long & \Ha, \NII \\
  CDF-S &  698 & 53.261 & -27.806 & 2.549 & 1++ & -99.0 & 3.101 & H-long & \Hb \\
  CDF-S &  699 &  53.261 & -27.760 & 1.257 & 2+ & 1.130 & 1.301 & H-short & \Ha, \NII, \SII \\
  CDF-S &  706 &  53.266 & -27.841 & 0.891 & 2 & 0.891 & 0.904 & J-long & \Ha, \NII, \SII \\
  CDF-S &  720 &  53.282 & -27.858 & 1.609 & 1 & 1.609 & 2.535 & J-long, H-long & \OIII, \Ha \\
  CDF-S &  724 &  53.287 & -27.694 & 1.337 & 2 & 1.335 & 1.321 & H-short & \Ha, \SII \\
  CDF-S &  728 &  53.292 & -27.812 & 1.013 & 2 & 1.034 & 1.050 & J-long, H-long & \Ha, \NII \\
  CDF-S &  731 & 53.302 & -27.776 & 0.610 & 1++ & -99.0 & 0.562 & J-long, H-short, H-long & \SII \\  
E-CDF-S &   61 &  52.870 & -27.983 & 0.752 & 2 & 0.752 & 0.763 & J-long & \Ha \\
E-CDF-S &   68 &  52.878 & -27.976 & 1.362 & 1 & 1.366 & 1.382 & H-short & \Ha, \NII \\
E-CDF-S &   85 &  52.889 & -27.805 & 0.822 & 1 & 0.678 & 0.669 & J-long, H-short & \Ha \\
E-CDF-S &   89 &  52.891 & -27.767 & 1.613 & 2++ & -99.0 & 1.889 & J-long, H-long & \Ha, \NII \\
E-CDF-S & 105 & 52.904 & -27.969 & 1.606 & 1++ & -99.0 & 1.368 & J-long, H-short, H-long & \OIII, \Ha, \SII \\
E-CDF-S &  115 &  52.911 & -27.996 & 1.190 & 2++ & -99.0 & 1.255 & J-long, H-short, H-long & \Ha, \NII \\
E-CDF-S &  118 & 52.916 & -27.699 & 0.469 & 2 & 0.467 & -99.0 & J-long & \SII \\
E-CDF-S &  121 &  52.917 & -27.940 & 5.137 & 2++ & -99.0 & 5.315 & J-long, H-long & \MgII \\
E-CDF-S &  137 &  52.930 & -28.020 & 1.197 & 2++ & -99.0 & 1.425 & J-long, H-short, H-long & \Ha, \NII, \SII \\
E-CDF-S &  150 &  52.935 & -27.943 & 0.727 & 2++ & -99.0 & 0.879 & J-long, H-long & \Ha, \SIII \\
E-CDF-S &  157 &  52.942 & -27.695 & 1.322 & 2+ & 1.315 & 1.184 & H-short & \Ha, \NII \\
E-CDF-S &  166 &  52.947 & -27.920 & 1.408 & 2 & 1.404 & 1.186 & H-short & \Ha, \NII \\
E-CDF-S &  193 & 52.963 & -27.954 & 1.161 & 1 & 1.167 & 1.153 & H-short & \Ha, \NII \\
E-CDF-S & 195 & 52.963 & -27.982 & 1.361 & 1 & 1.368 & 0.831 & H-short & \Ha, \NII \\
E-CDF-S &  201 &  52.966 & -27.947 & 1.618 & 2++ & -99.0 & 1.411 & J-long, H-short, H-long & \OIII, \Ha \\
E-CDF-S &  217 &  52.975 & -28.046 & 1.602 & 2++ & -99.0 & 1.578 & J-long, H-long & \OII, \OIII, \Ha, \NII \\
E-CDF-S &  218 & 52.976 & -27.997 & 0.738 & 1 & 0.740 & 0.731 & J-long & \Ha, \SII \\
E-CDF-S &  222 &  52.978 & -28.017 & 0.662 & 2+ & 0.623 & 0.677 & H-short & \SIII \\
E-CDF-S &  234 &  52.987 & -28.080 & 1.675 & 1++ & -99.0 & -99.0 & J-long, H-short, H-long & \Ha \\
E-CDF-S &  235 &  52.987 & -28.030 & 1.386 & 2 & 1.380 & 1.889 & H-short & \Ha, \NII \\
E-CDF-S &  276 &  53.018 & -28.075 & 1.295 & 2++ & -99.0 & -99.0 & J-long, H-short, H-long & \OIII, \Ha, \NII \\
E-CDF-S &  282 &  53.022 & -28.071 & 1.215 & 2++ & -99.0 & -99.0 & H-short & \Ha, \NII \\
E-CDF-S &  298 &  53.029 & -27.971 & 0.847 & 1++ & -99.0 & 1.067 & J-long, H-long & \SIII \\
E-CDF-S &  342 &  53.066 & -28.025 & 1.641 & 2++ & -99.0 & 1.651 & H-long & \Ha \\
E-CDF-S &  358 &  53.085 & -28.037 & 1.626 & 2 & 1.624 & 1.635 &  J-long, H-long & \Hb, \OIII, \Ha, \NII, \SII \\
E-CDF-S &  372 &  53.108 & -28.013 & 2.216 & 2++ & -99.0 & 2.052 & J-long, H-short, H-long & \Hb \\
E-CDF-S &  384 &  53.116 & -28.079 & 1.531 & 1++ & -99.0 & -99.0 & J-long, H-short, H-long & \Ha \\
E-CDF-S &  388 &  53.122 & -28.029 & 1.555 & 2+ & 0.638 & 1.591 & J-long, H-short, H-long & \Ha, \NII, \SII \\
E-CDF-S &  400 &  53.135 & -28.058 & 1.222 & 2 & 1.220 & 1.102 & J-long, H-short, H-long & \Ha, \NII \\
E-CDF-S &  411 & 53.151 & -27.589 & 1.226 & 2 & 1.220 & 1.199 & J-long, H-short & \Ha, \NII \\
E-CDF-S &  424 &  53.165 & -28.014 & 1.517 & 2++ & -99.0 & 1.507 & J-long, H-short, H-long & \Ha, \NII \\
E-CDF-S &  459 &  53.194 & -27.995 & 1.667 & 1++ & -99.0 & -99.0 & J-long, H-short, H-long & \Ha \\
E-CDF-S &  470 &  53.205 & -28.063 & 0.680 & 1 & 0.680 & 0.644 & J-long, H-short, H-long & \SIII \\
E-CDF-S &  481 &  53.213 & -28.034 & 2.529 & 1++ & -99.0 & 2.695 & J-long, H-short, H-long & \OIII \\
E-CDF-S &  517 &  53.247 & -27.603 & 1.345 & 2 & 1.350 & 1.337 & H-short & \Ha \\
E-CDF-S &  521 &  53.248 & -27.624 & 1.608 & 1++ & -99.0 & 1.484 & J-long, H-long & \Ha \\
E-CDF-S &  538 &  53.256 & -28.048 & 1.373 & 2+ & 0.200 & 1.812 & J-long, H-short, H-long & \Ha, \NII, \SII \\
E-CDF-S &  546 &  53.264 & -27.885 & 0.892 & 1+ & 1.346 & 0.954 & J-long, H-short & \Ha, \SII \\
E-CDF-S &  557 &  53.271 & -27.674 & 0.311 & 2 & 0.311 & 0.285 & J-long, H-short, H-long & \SIII \\
E-CDF-S &  571 &  53.278 & -27.774 & 1.705 & 2 & 1.705 & 1.497 & J-long, H-long & \Ha, \NII \\
E-CDF-S &  588 &  53.287 & -27.974 & 2.579 & 2 & 2.583 & 1.104 & H-long & \OIII \\
E-CDF-S &  595 &  53.290 & -27.736 & 1.037 & 1++ & -99.0 & 1.016 & J-long, H-long & \Ha, \NII \\
E-CDF-S &  601 &  53.294 & -27.963 & 1.598 & 2 & 1.598 & 1.726 & J-long, H-long & \Ha, \NII, \SII \\
E-CDF-S &  621 &  53.310 & -27.975 & 1.320 & 2++ & -99.0 & 1.555 & J-long, H-short, H-long & \OIII, \Ha, \NII \\
E-CDF-S &  624 & 53.311 & -27.706 & 1.320 & 1++ & -99.0 & 1.440 & J-long, H-short, H-long & \Ha \\
E-CDF-S &  635 &  53.319 & -27.952 & 1.045 & 2+ & 1.044 & 1.020 & J-long & \Ha \\
E-CDF-S &  678 &  53.345 & -27.923 & 1.629 & 2 & 1.628 & 1.633 & J-long, H-long & \Hb, \OIII, \Ha, \NII \\
E-CDF-S &  681 &  53.347 & -27.771 & 0.834 & 2 & 0.835 & 0.809 & J-long & \Ha \\
E-CDF-S &  688 &  53.349 & -27.725 & 0.787 & 2 & 0.787 & 0.786 & J-long & \Ha, \NII \\
E-CDF-S &  698 &  53.357 & -27.816 & 1.413 & 1++ & -99.0 & 1.464 & J-long, H-short, H-long & \Ha, \NII \\
E-CDF-S &  701 &  53.360 & -27.722 & 0.964 & 2 & 0.964 & 0.950 & J-long & \Ha, \NII \\
E-CDF-S &  705 & 53.362 & -27.755 & 1.658 & 1++ & -99.0 & 1.835 & H-long & \Ha, \NII, \SII \\
E-CDF-S &  709 &  53.368 & -27.972 & 1.558 & 1++ & -99.0 & -99.0 & J-long, H-short & \OIII \\
E-CDF-S &  715 &  53.371 & -27.732 & 1.280 & 1++ & -99.0 & -99.0 & J-long, H-short & \Ha \\
E-CDF-S &  716 &  53.372 & -27.991 & 0.763 & 2 & 0.763 & 0.768 & J-long & \Ha, \NII, \SII \\
E-CDF-S &  718 & 53.374 & -27.852 & 0.946 & 1 & 0.951 & 0.956 & H-short, H-long & \SII \\
E-CDF-S &  725 &  53.383 & -27.903 & 1.314 & 2 & 1.315 & -99.0 & J-long, H-short, H-long & \Ha, \NII \\
E-CDF-S &  728 &  53.387 & -27.819 & 1.583 & 2 & 1.581 & 1.568 & J-long, H-short, H-long & \Ha, \NII, \SII \\
 XMM-LH &    5 & 163.180 &  57.263 & 2.138 & 2 & 2.144 & 2.050 & H-short, H-long & \Hb, \OIII \\
 XMM-LH &   15 & 163.184 &  57.286 & 2.010 & 1++ & -99.0 & 2.022 & H-short & \OIII \\
 XMM-LH &   19 &  163.240 & 57.292 & 0.746 & 2++ & -99.0 & 0.730 & H-short, H-long & \SIII \\
 XMM-LH &   25 & 163.348 &  57.293 & 1.599 & 2++ & -99.0 & 1.450 & H-short, H-long & \Ha, \NII \\
 XMM-LH &   34 & 163.253 &  57.303 & 0.517 & 2++ & -99.0 & 0.520 & H-long & He\,\textsc{i} \\
 XMM-LH &   35 & 163.470 &  57.304 & 1.727 & 2++ & -99.0 & 1.441 & H-short, H-long & \Ha, \NII \\
 XMM-LH &   48 & 163.250 &  57.317 & 1.458 & 1++ & -99.0 & 1.460 & H-short, H-long & \Ha \\
 XMM-LH &   50 & 163.107 &  57.318 & 4.470 & 2 & 4.449 & 4.340 & J-long, H-short, H-long & \MgII \\
 XMM-LH &   51 & 163.032 &  57.318 & 2.677 & 1++ & -99.0 & 2.760 & H-long & \Hb, \OIII \\
 XMM-LH &   61 & 162.990 &  57.328 & 1.378 & 2+ & 1.379 & 1.188 & H-short & \Ha, \NII, \SII \\
 XMM-LH &   63 & 163.151 &  57.267 & 1.729 & 2 & 1.734 & 1.649 & J-long, H-long & \Hb, \Ha, \NII \\
 XMM-LH &   65 & 163.231 &  57.331 & 1.452 & 2 & 1.450 & 1.296 & J-long, H-short, H-long & \Ha, \NII, \SII \\
 XMM-LH &   72 & 163.070 &  57.338 & 2.741 &1 & 2.710 & 1.060 & H-long & \Hg, \OIII \\
 XMM-LH &   85 & 163.373 &  57.351 & 1.144 & 2 & 1.145 & 0.469 & J-long, H-short, H-long & \Ha, \NII, \SII \\
 XMM-LH &   96 & 163.187 &  57.356 & 2.826 & 1 & 2.832 & -99.0 & J-long, H-short, H-long & \Hg \\
 XMM-LH &  103 & 163.324 &  57.364 & 1.484 & 2++ & -99.0 & 1.494 & H-short, H-long & \Ha, \NII \\
 XMM-LH &  119 & 163.510 &  57.376 & 1.406 & 2++ & -99.0 & 2.030 & H-short, H-long & \Ha, \NII, \SII \\
 XMM-LH &  120 & 163.105 &  57.385 & 1.523 & 2 & 1.524 & 1.488 & J-long, H-long & \OIII, \Ha, \NII, \SII \\
 XMM-LH &  121 & 163.243 &  57.381 & 0.763 & 2 & 0.762 & 0.520 & J-long, H-short, H-long & \Ha, \NII, \SII \\
 XMM-LH &  124 & 163.299 &  57.385 & 1.534 & 2 & 1.552 & -99.0 & H-long & \Ha \\
 XMM-LH &  131 & 163.173 &  57.389 & 1.005 & 1 & 1.013 & 1.107 & J-long, H-long & \Ha \\
 XMM-LH &  136 & 163.050 & 57.389 & 0.743 & 1++ & -99.0 & 1.170 & H-short, H-long & \SIII \\
 XMM-LH &  151 & 163.483 &  57.399 & 1.398 & 2++ & -99.0 & 1.445 & H-short, H-long & \Ha, \NII \\
 XMM-LH &  154 & 163.609 &  57.401 & 0.963 & 2+ & 0.963 & 1.049 & J-long & \Ha \\
 XMM-LH &  156 & 162.977 & 57.402 & 2.391 & 2 & 2.367 & 0.544 & J-long, H-short, H-long & \Hb \\
 XMM-LH &  167 & 163.574 &  57.406 & 1.220 & 1++ & -99.0 & 1.406 & H-short & \Ha \\
 XMM-LH &  168 & 163.383 &  57.415 & 1.958 & 2 & 1.956 & 1.921 & J-long, H-short, H-long & \Hb, \OIII \\
 XMM-LH &  171 & 163.588 &  57.429 & 0.205 & 2 & 0.205 & 0.214 & J-long & \SIII, He\,\textsc{i} \\
 XMM-LH &  174 & 162.837 &  57.416 & 1.186 & 1++ & -99.0 & 1.168 & H-short & \Ha \\
 XMM-LH &  176 & 163.239 &  57.419 & 1.533 & 2 & 1.527 & 0.501 & J-long, H-long & \Hb, \Ha, \NII \\
 XMM-LH &  179 & 163.132 & 57.417 & 1.475 & 1++ & -99.0 & 1.480 & H-short, H-long & \Ha, \NII \\
 XMM-LH &  183 & 163.301 &  57.418 & 1.877 & 2 & 1.876 & 1.603 & J-long, H-short, H-long & \OIII \\
 XMM-LH &  191 & 163.396 &  57.428 & 0.787 & 2++ & -99.0 & 0.967 & J-long, H-short, H-long & \Ha, \NII, \SII \\
 XMM-LH &  203 & 163.500 &  57.434 & 1.742 & 2++ & -99.0 & 1.772 & H-long & \Ha, \OI \\
 XMM-LH &  217 & 162.799 &  57.443 & 0.760 & 1 & 0.758 & 0.734 & J-long, H-long & \Ha \\
 XMM-LH &  229 & 162.935 &  57.447 & 1.285 & 2++ & -99.0 & 1.289 & H-short, H-long & \Ha, \NII \\
 XMM-LH &  247 & 163.358 &  57.461 & 1.367 & 1++ & -99.0 & 1.379 & H-short & \Ha \\
 XMM-LH &  254 & 163.181 &  57.466 & 1.183 & 1 & 1.210 & 0.968 & J-long, H-short, H-long & \Ha \\
 XMM-LH &  261 & 162.937 &  57.469 & 3.406 & 2 & 3.408 & 3.022 & J-long, H-long & \MgII, \OII \\
 XMM-LH &  267 & 163.352 &  57.472 & 1.549 & 2 & 1.563 & 1.420 & J-long, H-long & \OIII, \Ha, \NII \\
 XMM-LH &  268 & 163.450 &  57.471 & 1.295 & 1++ & -99.0 & 1.298 & H-short & \Ha \\
 XMM-LH &  270 & 163.289 &  57.472 & 1.576 & 2 & 1.575 & 1.049 & J-long, H-long & \Hb, \Ha, \SII \\
 XMM-LH &  277 & 163.437 & 57.478 & 1.807 & 1 & 1.816 & 1.945 & J-long, H-long & \Hg \\ 
 XMM-LH &  279 & 163.192 & 57.472 & 1.514 & 1++ & -99.0 & 1.460 & H-short, H-long & \Ha, \NII \\ 
 XMM-LH &  300 & 163.266 &  57.490 & 0.789 & 2 & 0.788 & 0.504 & J-long, H-short, H-long & \Ha, \NII, \SII, \SIII \\
 XMM-LH &  306 & 163.028 &  57.490 & 0.709 & 1 & 0.709 & 0.677 & J-long, H-long & \SIII \\
 XMM-LH &  321 & 163.102 &  57.502 & 1.008 & 2 & 1.009 & 1.543 & J-long, H0long & \Ha, \NII \\
 XMM-LH &  340 & 163.130 &  57.504 & 1.212 & 2 & 1.212 & 1.034 & H-short & \Ha, \NII, \SII \\
 XMM-LH &  342 & 163.416 &  57.518 & 0.587 & 2 & 0.586 & 0.533 & J-long, H-short, H-long & \SIII, He\,\textsc{i} \\
 XMM-LH &  354 & 162.856 &  57.514 & 3.409 & 2 & 3.409 & 2.945 & J-long, H-short, H-long & \MgII, \OII \\
 XMM-LH &  355 & 163.155 &  57.518 & 0.740 & 1 & 0.710 & 0.583 & J-long, H-long & \Ha, \SIII \\
 XMM-LH &  359 & 162.980 &  57.512 & 1.102 & 1++ & -99.0 & 1.087 & H-short & \SII \\
 XMM-LH & 370 & 163.442 & 57.518 & 1.650 & 1++ & -99.0 & 1.640 & H-short, H-long & \Ha, \NII \\
 XMM-LH &  385 & 163.176 &  57.533 & 1.379 & 2 & 1.379 & 1.288 & H-short & \Ha, \NII \\
 XMM-LH &  386 & 163.058 &  57.528 & 0.894 & 2 & 0.896 & 0.849 & J-long & \Ha, \SII \\
 XMM-LH &  387 & 163.249 &  57.532 & 1.449 & 2 & 1.447 & 1.398 & H-short, H-long & \Ha \\
 XMM-LH &  406 & 163.488 &  57.545 & 1.283 & 2 & 1.296 & 1.721 & J-long, H-short & \Ha \\
 XMM-LH &  409 & 163.403 &  57.550 & 1.600 & 2 & 1.601 & 0.504 & J-long, H-long & \Ha, \NII \\
 XMM-LH &  419 & 163.501 &  57.556 & 1.491 & 1++ & -99.0 & 1.491 & H-long & \Ha \\
 XMM-LH &  429 & 163.559 &  57.558 & 2.913 & 1++ & -99.0 & 2.789 & H-short, H-long & \OII \\
 XMM-LH &  430 & 162.785 &  57.562 & 1.553 & 2 & 1.539 & 1.665 & H-short, H-long & \Ha \\
 XMM-LH &  443 & 163.155 &  57.567 & 1.877 & 2 & 1.877 & 1.823 & J-long, H-short & \OIII \\
 XMM-LH &  450 & 163.275 &  57.573 & 2.952 & 2 & 2.945 & 2.529 & J-long, H-short, H-long & \OII, \Hg, \OIII \\
 XMM-LH &  453 & 163.302 &  57.573 & 1.214 & 2 & 1.213 & 1.748 & H-short & \Ha, \SII \\
 XMM-LH &  456 & 162.977 &  57.577 & 0.877 & 2 & 0.877 & 0.648  & J-long, H-short, H-long & \Ha, \NII, \SII, \SIII \\
 XMM-LH &  468 & 163.171 &  57.580 & 1.370 & 2++ & -99.0 & 1.374 & H-short & \Ha, \SII \\
 XMM-LH & 472 & 163.417 & 57.588 & 1.136 & 1 & 1.135 & 0.899 & H-short & \SII \\ 
 XMM-LH &  475 & 163.320 &  57.597 & 1.205 & 2++ & -99.0 & 1.219 & J-long, H-short, H-long & \Ha, \NII \\
 XMM-LH &  477 & 163.283 &  57.589 & 1.468 & 2++ & -99.0 & 0.777 & H-long & \Ha \\
 XMM-LH &  488 & 163.369 &  57.594 & 0.690 & 1 & 0.700 & 0.584 & H-long & \SIII \\
 XMM-LH &  518 & 163.401 &  57.624 & 2.269 & 2++ & -99.0 & 2.563 & H-short, H-long & \OIII \\
 XMM-LH &  521 & 163.176 &  57.621 & 1.591 & 1++ & -99.0 & 1.526 & H-long & \Ha \\
 XMM-LH &  523 & 162.875 &  57.627 & 1.217 & 2++ & -99.0 & 1.192 & H-short, H-long & \Ha, \NII \\
 XMM-LH &  527 & 163.260 &  57.632 & 1.885 & 1 & 1.881 & 1.748 & J-long, H-short & \Hb, \OIII \\
 XMM-LH &  528 & 163.151 &  57.628 & 2.341 & 1 & 2.352 & 2.076 & H-short & \Hg, \OIII \\
 XMM-LH &  529 & 162.905 &  57.632 & 1.940 & 2+ & 1.910 & 1.997 & J-long, H-short, H-long & \OIII \\
 XMM-LH &  530 & 162.844 &  57.626 & 1.474 & 2++ & -99.0 & 1.426 & H-short, H-long & \Ha, \SII \\
 XMM-LH &  532 & 163.190 &  57.629 & 1.675 & 2 & 1.677 & 0.534 & J-long, H-long & \Hb, \OIII, \Ha, \NII, \SII \\
 XMM-LH &  553 & 163.125 &  57.654 & 1.440 & 2 & 1.437 & 0.454 & J-long, H-short, H-long & \Hb, \OIII, \Ha, \NII, \SII \\
 XMM-LH &  555 & 162.967 &  57.652 & 1.674 & 2++ & -99.0 & 2.080 & H-short, H-long & \Ha, \NII \\
 XMM-LH & 560 & 163.216 & 57.652 & 1.724 & 1++ & -99.0 & 1.650 & H-long & \Ha, \NII \\  
 XMM-LH &  563 & 163.379 &  57.656 & 1.379 & 2++ & -99.0 & 1.559 & H-short, H-long & \Ha, \NII, \SII \\
 XMM-LH &  569 & 163.240 &  57.663 & 0.947 & 1++ & -99.0 & 1.044 & H-short, H-long & \SIII \\
 XMM-LH & 572 & 163.343 & 57.669 & 0.750 & 1++ & -99.0 & 0.730 & H-short, H-long & \SIII \\
 XMM-LH &  588 & 163.068 &  57.686 & 1.409 & 2++ & -99.0 & 1.272 & H-short, H-long & \Ha \\
 XMM-LH &  591 & 163.097 &  57.689 & 1.535 & 2 & 1.534 & 1.527 & J-long, H-short, H-long & \OIII, \Ha, \NII, \SII \\
 XMM-LH &  594 & 163.202 &  57.691 & 0.814 & 2+ & 0.814 & 1.087 & J-long, H-long & \Ha, \NII \\
 XMM-LH &  595 & 162.932 &  57.689 & 1.602 & 2++ & -99.0 & 1.583 & H-long & \Ha \\
 XMM-LH &  599 & 163.280 &  57.251 & 2.416 & 2 & 2.416 & 2.159 & J-long, H-short, H-long & \OII, \Hb, \OIII \\
 XMM-LH &  604 & 163.299 &  57.704 & 2.104 & 2 & 2.113 & 1.932 & J-long, H-short, H-long & \Hb, \OIII \\
 XMM-LH &  610 & 163.166 &  57.719 & 0.679 & 1+ & 0.679 & 0.675 & J-long, H-short, H-long & \SII \\
 XMM-LH &  999 & 163.105 &  57.380 & 0.809 & 2 & 0.807 & 0.659 & J-long, H-short, H-long & \Ha, \NII, \SII \\
 XMM-LH & 1379 & 163.559 &  57.462 & 1.689 & 1++ & -99.0 & 1.440 & H-long & \Ha \\
 XMM-LH & 1395 & 162.838 &  57.327 & 1.649 & 2++ & -99.0 & 1.672 & H-short, H-long & \Ha \\
 XMM-LH & 1443 & 163.141 &  57.279 & 1.391 & 2++ & -99.0 & 1.415 & H-short & \Ha \\
 XMM-LH & 1458 & 162.855 &  57.484 & 0.484 & 1 & 0.484 & 0.502 & H-short & He\,\textsc{i} \\
 XMM-LH & 1476 & 163.360 & 57.651 & 1.235 & 1++ & -99.0 & 1.300 & H-short & \Ha, \NII \\ 
 XMM-LH & 1518 & 163.525 &  57.404 & 0.775 & 2++ & -99.0 & 0.824 & J-long, H-long & \Ha \\
 XMM-LH & 1551 & 163.461 &  57.459 & 1.673 & 2++ & -99.0 & 1.704 & H-short, H-long & \Ha, \NII \\
 XMM-LH & 1595 & 163.071 & 57.610 & 1.545 & 1++ & -99.0 & 1.580 & H-long & \Ha \\
 XMM-LH & 2015 & 163.258 & 57.508 & 1.357 & 1++ & -99.0 & 1.370 & H-short & \Ha, \NII, \SII \\
 XMM-LH & 2020 & 163.459 &  57.452 & 1.728 & 2+ & 1.720 & 1.442 & J-long, H-long & \Ha \\
 XMM-LH & 2081 & 163.035 &  57.346 & 1.356 & 1++ & -99.0 & 1.296 & H-short & \Ha, \NII, \SII \\
 XMM-LH & 2084 & 163.236 &  57.400 & 1.524 & 2++ & -99.0 & 1.348 & H-short, H-long & \Ha, \NII, \SII \\
 XMM-LH & 2254 & 163.123 &  57.726 & 2.945 & 2++ & -99.0 & 2.803 & H-short, H-long & \Hg \\
 XMM-LH & 2278 & 163.016 &  57.452 & 2.438 & 2 & 2.454 & 2.083 & J-long, H-short, H-long & \Hb, \OIII \\
 XMM-LH & 2374 & 163.460 &  57.500 & 2.508 & 2++ & -99.0 & -99.0 & H-long & \OIII \\
 XMM-LH & 2379 & 163.181 & 57.418 & 0.752 & 1++ & -99.0 & 0.750 & H-long & \SIII \\ 
 XMM-LH & 2380 & 163.423 &  57.507 & 2.063 & 1++ & -99.0 & 2.095 & H-short, H-long & \Hb, \OIII \\
 XMM-LH & 2506 & 162.890 &  57.590 & 2.010 & 2++ & -99.0 & 2.016 & J-long, H-short & \OIII \\
\end{longtable*}

\begin{longtable*}{l r r r r r r r r r r}
  \caption{Emission line properties of Broad-line AGNs} \\
 \hline \hline
    \multicolumn{1}{c}{\textbf{Field}} & 
    \multicolumn{1}{r}{\textbf{ID}} & 
    \multicolumn{1}{c}{\textbf{$z$}} & 
    \multicolumn{1}{r}{\textbf{log L$_{\rm bol}$}} & 
    \multicolumn{1}{r}{\textbf{log M$_{\rm BH}$}} & 
    \multicolumn{3}{c}{\textbf{log FWHM (km s$^{-1}$)}} &
    \multicolumn{3}{c}{\textbf{log L (erg s$^{-1}$)}}  \\
    \multicolumn{1}{c}{} &
    \multicolumn{1}{c}{} &
    \multicolumn{1}{c}{} &
    \multicolumn{1}{c}{\textbf{(erg s$^{-1}$)}} &
    \multicolumn{1}{c}{\textbf{(M$_{\odot}$)}} &
    \multicolumn{1}{c}{\textbf{\Ha}} &
    \multicolumn{1}{c}{\textbf{\Hb}} &
    \multicolumn{1}{c}{\textbf{MgII}} &
    \multicolumn{1}{c}{\textbf{\Ha}} &
    \multicolumn{1}{c}{\textbf{\Hb}} &
    \multicolumn{1}{c}{\textbf{3000\AA}} \\  \hline \\[-1.8ex]
\endfirsthead
\multicolumn{6}{c}{\tablename \thetable{} -- continued.} \\[0.5ex]
\hline \hline \\[-1.8ex]
    \multicolumn{1}{c}{\textbf{Field}} & 
    \multicolumn{1}{r}{\textbf{ID}} & 
    \multicolumn{1}{c}{\textbf{$z$}} & 
    \multicolumn{1}{r}{\textbf{log L$_{\rm bol}$}} & 
    \multicolumn{1}{r}{\textbf{log M$_{\rm BH}$}} & 
    \multicolumn{3}{c}{\textbf{log FWHM (km s$^{-1}$)}} &
    \multicolumn{3}{c}{\textbf{log L (erg s$^{-1}$)}}  \\
    \multicolumn{1}{c}{} &
    \multicolumn{1}{c}{} &
    \multicolumn{1}{c}{} &
    \multicolumn{1}{c}{\textbf{(erg s$^{-1}$)}} &
    \multicolumn{1}{c}{\textbf{(M$_{\odot}$)}} &
    \multicolumn{1}{c}{\textbf{\Ha}} &
    \multicolumn{1}{c}{\textbf{\Hb}} &
    \multicolumn{1}{c}{\textbf{MgII}} &
    \multicolumn{1}{c}{\textbf{\Ha}} &
    \multicolumn{1}{c}{\textbf{\Hb}} &
    \multicolumn{1}{c}{\textbf{3000\AA}} \\  \hline \\[-1.8ex]
    \endhead
\\[-1.8ex] \hline \hline
\endfoot
  CDF-S &    1 & 1.630 & 45.89$\pm$0.07 & 8.41$\pm$0.08 & 3.74$\pm$0.02 & \nodata &  \nodata & 43.08$\pm$0.06 &  \nodata  &  \nodata  \\
  CDF-S &    4 & 1.270 & 45.39$\pm$0.14 & 7.22$\pm$0.24 & 3.40$\pm$0.05 &  \nodata &  \nodata  & 42.16$\pm$0.12 &  \nodata  &  \nodata \\
  CDF-S &   11 & 1.888 & 46.02$\pm$0.06 & 8.37$\pm$0.16 &  \nodata &  \nodata & 3.61$\pm$0.02 &  \nodata &  \nodata  & 45.52$\pm$0.00 \\
  CDF-S &   14 & 1.370 & 44.32$\pm$0.05 & 8.29$\pm$0.05 & \nodata  &  \nodata & 3.68$\pm$0.13 &  \nodata &  \nodata  & 45.19$\pm$0.01 \\
  CDF-S &   15 & 1.065 & 44.25$\pm$0.04 & 6.52$\pm$0.07 &  \nodata &  \nodata & 3.36$\pm$0.33 &  \nodata &  \nodata  & 43.34$\pm$0.03 \\
  CDF-S &   25 & 1.336 & 44.34$\pm$0.13 & 6.69$\pm$0.09 & \nodata & \nodata & 3.32$\pm$0.04 &  \nodata &  \nodata & 43.51$\pm$0.02 \\
  CDF-S &   66 & 0.575 & 42.80$\pm$0.06 & 7.67$\pm$0.58 &  \nodata & 3.39$\pm$0.60 &  \nodata &  \nodata & 41.65$\pm$0.00 &  \nodata \\
  CDF-S &   76 & 1.042 & 45.06$\pm$0.02 & 7.45$\pm$0.07 & 3.63$\pm$0.08 &  \nodata & 3.68$\pm$0.02 & 42.04$\pm$0.17 &   & 43.81$\pm$0.00 \\
  CDF-S &   87 & 1.437 & 44.15$\pm$0.06 & 8.67$\pm$0.06 &  \nodata & \nodata  & 4.41$\pm$0.17 &  \nodata  &  \nodata  & 43.43$\pm$0.02 \\
  CDF-S &   88 & 1.613 & 45.14$\pm$0.02 & 7.05$\pm$0.02 & \nodata  & \nodata  & 3.67$\pm$0.01 &  \nodata &  \nodata  & 43.22$\pm$0.01 \\
  CDF-S &  101 & 0.966 & 45.36$\pm$0.05 & 7.40$\pm$0.16 & 3.70$\pm$0.05 &  \nodata & 3.73$\pm$0.02 & 42.23$\pm$0.11 &  \nodata  & 43.57$\pm$0.00 \\
  CDF-S &  166 & 1.608 & 45.67$\pm$0.01 & 8.30$\pm$0.19 & 3.64$\pm$0.07 &  \nodata & 3.84$\pm$0.16 & 42.64$\pm$0.19 &  \nodata  & 44.68$\pm$0.00 \\
  CDF-S &  229 & 1.326 & 45.68$\pm$0.01 & 7.75$\pm$0.02 & 3.35$\pm$0.33 & 3.37$\pm$0.05 & 3.71$\pm$0.14 & 43.30$\pm$0.00 & 41.93$\pm$0.14 & 45.26$\pm$0.01 \\
  CDF-S &  241 & 0.566 & 44.07$\pm$0.02 & 7.26$\pm$0.06 &  \nodata &  \nodata & 3.67$\pm$0.14 &  \nodata &  \nodata  & 43.55$\pm$0.01 \\
  CDF-S &  329 & 0.954 & 44.26$\pm$0.13 & 7.03$\pm$0.86 & 3.39$\pm$0.09 &  \nodata &  \nodata & 41.86$\pm$0.27 &  \nodata  &  \nodata \\
  CDF-S &  344 & 1.615 & 45.03$\pm$0.02 & 8.06$\pm$0.09 & 3.35$\pm$0.23 &  \nodata & 3.89$\pm$0.01 & 43.02$\pm$0.16 &  \nodata  & 44.13$\pm$0.02 \\
  CDF-S &  367 & 1.041 & 45.45$\pm$0.00 & 7.27$\pm$0.38 &  \nodata &  \nodata & 3.39$\pm$0.02 &  \nodata &  \nodata  & 44.46$\pm$0.00 \\
  CDF-S &  369 & 1.612 & 45.53$\pm$0.02 & 7.46$\pm$0.82 & 3.47$\pm$0.16 &  \nodata &  \nodata & 42.35$\pm$0.15 &  \nodata  &  \nodata \\
  CDF-S &  375 & 0.742 & 45.76$\pm$0.00 & 8.14$\pm$0.02 & 3.78$\pm$0.24 &  \nodata & 3.81$\pm$0.01 & 43.12$\pm$0.03 &  \nodata  & 44.51$\pm$0.01 \\
  CDF-S &  417 & 1.222 & 45.33$\pm$0.01 & 8.91$\pm$0.14 & 4.03$\pm$0.15 &  \nodata &  \nodata & 42.90$\pm$0.12 &  \nodata  &  \nodata \\
  CDF-S &  420 & 0.960 & 44.09$\pm$0.03 & 7.35$\pm$0.30 & 3.38$\pm$0.65 &  \nodata &  \nodata & 42.47$\pm$0.00 &  \nodata  &  \nodata \\
  CDF-S &  473 & 1.557 & 45.82$\pm$0.02 & 7.84$\pm$0.52 & 3.50$\pm$0.59 &  \nodata &  \nodata & 42.92$\pm$0.00 &  \nodata  &  \nodata \\
  CDF-S &  514 & 0.664 & 44.03$\pm$0.03 & 7.46$\pm$0.61 &  \nodata &  & 3.85$\pm$0.27 &  \nodata &  \nodata  & 43.27$\pm$0.00 \\
  CDF-S &  523 & 0.838 & 45.43$\pm$0.01 & 8.35$\pm$0.03 & 3.69$\pm$0.40 &  \nodata & 4.11$\pm$0.29 & 43.14$\pm$0.00 &  \nodata  & 44.56$\pm$0.03 \\
  CDF-S &  537 & 1.216 & 44.87$\pm$0.02 & 8.60$\pm$0.16 & \nodata  &  \nodata & 3.96$\pm$0.17 &  \nodata &  \nodata  & 44.76$\pm$0.02 \\
  CDF-S &  614 & 0.664 & 43.18$\pm$0.06 & 8.23$\pm$0.03 &  \nodata &  \nodata & 4.14$\pm$0.02 &  \nodata &  \nodata  & 43.58$\pm$0.02 \\
  CDF-S &  627 & 0.736 & 44.80$\pm$0.01 & 7.29$\pm$0.12 &  \nodata & \nodata  & 3.72$\pm$0.02 &  \nodata &  \nodata  & 43.43$\pm$0.02 \\
  CDF-S &  656 & 1.367 & 43.66$\pm$0.07 & 7.45$\pm$0.49 & 3.36$\pm$0.49 &  \nodata &  \nodata & 42.72$\pm$0.00 &  \nodata  &  \nodata \\
  CDF-S &  681 & 0.733 & 44.34$\pm$0.02 & 7.28$\pm$0.11 &  \nodata &  \nodata & 3.65$\pm$0.11 &  \nodata &  \nodata  & 43.65$\pm$0.00 \\
  CDF-S &  691 & 2.005 & 45.93$\pm$0.05 & 8.95$\pm$0.39 &  \nodata & \nodata  & 3.81$\pm$0.15 &  \nodata &  \nodata  & 45.82$\pm$0.01 \\
  CDF-S &  695 & 0.622 & 43.79$\pm$0.03 & 7.65$\pm$0.05 &  \nodata & \nodata  & 3.87$\pm$0.04 &  \nodata &  \nodata  & 43.51$\pm$0.02 \\
  CDF-S &  720 & 1.609 & 45.94$\pm$0.02 & 8.50$\pm$0.12 & 3.85$\pm$0.04 &  \nodata &  \nodata & 42.81$\pm$0.08 &  \nodata  &  \nodata \\
  CDF-S &  723 & 2.072 & 45.66$\pm$0.05 & 8.82$\pm$0.02 & \nodata  & \nodata  & 3.82$\pm$0.01 &  \nodata &  \nodata  & 45.55$\pm$0.02 \\
  CDF-S &  724 & 1.337 & 44.95$\pm$0.07 & 7.69$\pm$0.16 & 3.52$\pm$0.30 &  \nodata &  \nodata & 42.81$\pm$0.00 &  \nodata  &  \nodata \\
E-CDF-S &    7 & 1.368 & 46.16$\pm$0.00 & 9.30$\pm$0.04 & 4.12$\pm$0.02 &  \nodata & 3.90$\pm$0.00 & 43.27$\pm$0.04 &  \nodata  & 45.29$\pm$0.02 \\
E-CDF-S &   53 & 1.524 & 45.58$\pm$0.00 & 9.18$\pm$0.19 &  \nodata &  \nodata & 4.03$\pm$0.18 &  \nodata &  \nodata  & 45.48$\pm$0.02 \\
E-CDF-S &   68 & 1.362 & 44.58$\pm$0.03 & 8.55$\pm$0.06 &  \nodata &  \nodata & 3.77$\pm$0.03 &  \nodata &  \nodata  & 45.30$\pm$0.02 \\
E-CDF-S &   89 & 1.613 & 44.44$\pm$0.16 & 7.32$\pm$0.14 & 3.44$\pm$0.08 & \nodata  &  \nodata & 42.22$\pm$0.17 & \nodata   &  \nodata \\
E-CDF-S &  100 & 1.957 & 45.64$\pm$0.01 & 8.69$\pm$0.12 & \nodata  & \nodata  & 3.73$\pm$0.13 & \nodata  &  \nodata  & 45.65$\pm$0.01 \\
E-CDF-S &  158 & 0.717 & 44.44$\pm$0.01 & 8.81$\pm$0.08 &  \nodata & 3.89$\pm$0.02 &  \nodata &  \nodata & 41.88$\pm$0.03 &  \nodata \\
E-CDF-S &  166 & 1.408 & 44.81$\pm$0.02 & 8.31$\pm$0.61 & 3.70$\pm$0.42 & \nodata  &  \nodata & 43.02$\pm$0.00 &  \nodata  & \nodata  \\
E-CDF-S &  193 & 1.167 & 44.54$\pm$0.01 & 8.45$\pm$0.04 &  \nodata &  \nodata & 3.83$\pm$0.02 &  \nodata &  \nodata  & 44.96$\pm$0.05 \\
E-CDF-S &  358 & 1.626 & 45.64$\pm$0.00 & 8.29$\pm$0.02 & 3.42$\pm$0.43 & 3.73$\pm$0.09 & 3.61$\pm$0.00 & 43.59$\pm$0.00 & 41.95$\pm$0.21 & 45.38$\pm$0.01 \\
E-CDF-S &  381 & 0.526 & 44.28$\pm$0.00 & 7.72$\pm$0.34 & \nodata  & 3.63$\pm$0.17 & \nodata  & \nodata  & 40.87$\pm$0.19 &  \nodata \\
E-CDF-S &  517 & 1.345 & 44.94$\pm$0.00 & 7.82$\pm$0.30 & 3.64$\pm$0.08 &   \nodata &  \nodata & 42.37$\pm$0.23 &  \nodata  &   \nodata\\
E-CDF-S &  601 & 1.598 & 45.47$\pm$0.02 & 7.55$\pm$0.35 & 3.41$\pm$0.59 &  \nodata &  \nodata & 42.74$\pm$0.00 &  \nodata  &  \nodata \\
E-CDF-S &  631 & 2.072 & 45.49$\pm$0.01 & 8.67$\pm$0.24 &  \nodata &  \nodata & 3.79$\pm$0.06 &  \nodata &   \nodata & 45.42$\pm$0.05 \\
E-CDF-S &  678 & 1.629 & 45.30$\pm$0.00 & 8.58$\pm$0.06 & 3.77$\pm$0.18 &  \nodata & 3.73$\pm$0.21 & 43.25$\pm$0.07 &  \nodata  & 45.43$\pm$0.01 \\
E-CDF-S &  681 & 0.834 & 44.66$\pm$0.00 & 8.41$\pm$0.06 & 3.88$\pm$0.02 &  \nodata &  \nodata & 42.53$\pm$0.04 &  \nodata  &  \nodata \\
E-CDF-S &  700 & 2.171 & 45.78$\pm$0.00 & 8.31$\pm$0.24 &  \nodata &  \nodata & 3.91$\pm$0.19 &  \nodata &  \nodata  & 44.45$\pm$0.02 \\
E-CDF-S &  712 & 0.841 & 45.74$\pm$0.00 & 8.00$\pm$0.02 & 3.76$\pm$0.34 &  \nodata & 3.73$\pm$0.01 & 42.66$\pm$0.00 &  \nodata  & 44.55$\pm$0.07 \\
E-CDF-S &  716 & 0.763 & 44.93$\pm$0.00 & 8.11$\pm$0.43 & 3.70$\pm$0.33 &  \nodata &  \nodata & 42.65$\pm$0.00 &  \nodata  &   \nodata \\
E-CDF-S &  725 & 1.314 & 45.74$\pm$0.00 & 8.31$\pm$0.10 & 3.59$\pm$0.23 &  \nodata & 3.71$\pm$0.01 & 43.11$\pm$0.13 &  \nodata  & 45.12$\pm$0.02 \\
E-CDF-S &  728 & 1.583 & 45.44$\pm$0.01 & 8.54$\pm$0.19 & 3.81$\pm$0.03 &  \nodata &  \nodata & 43.02$\pm$0.05 &  \nodata  &  \nodata \\
E-CDF-S &  742 & 1.762 & 44.85$\pm$0.03 & 8.33$\pm$0.18 &  \nodata &  \nodata & 3.60$\pm$0.03 &  \nodata &  \nodata  & 45.48$\pm$0.01 \\
 XMM-LH &    5 & 2.138 & 46.55$\pm$0.01 & 8.24$\pm$0.08 &  \nodata & 3.35$\pm$0.54 &  \nodata &  \nodata & 42.77$\pm$0.00 &  \nodata \\
 XMM-LH &   25 & 1.599 & 44.52$\pm$0.32 & 7.52$\pm$0.65 & 3.50$\pm$0.07 & \nodata  &  \nodata & 42.36$\pm$0.17 &  \nodata  &  \nodata \\
 XMM-LH &   41 & 1.653 & 45.89$\pm$0.08 & 8.09$\pm$0.20 &  \nodata &  \nodata & 3.71$\pm$0.15 &  \nodata &  \nodata  & 44.75$\pm$0.02 \\
 XMM-LH &   85 & 1.144 & 45.26$\pm$0.01 & 8.03$\pm$0.02 &  \nodata &  \nodata & 3.75$\pm$0.18 &  \nodata &  \nodata  & 44.52$\pm$0.02 \\
 XMM-LH &  119 & 1.406 & 45.25$\pm$0.07 & 8.55$\pm$0.08 & 3.83$\pm$0.02 &  \nodata &  \nodata & 42.97$\pm$0.03 &  \nodata  &  \nodata \\
 XMM-LH &  120 & 1.523 & 45.65$\pm$0.02 & 7.62$\pm$0.16 & 3.54$\pm$0.09 &  \nodata & \nodata  & 42.39$\pm$0.25 &  \nodata  &  \nodata \\
 XMM-LH &  148 & 1.116 & 46.58$\pm$0.00 & 8.93$\pm$0.06 &  \nodata &  \nodata & 3.74$\pm$0.13 &  \nodata &  \nodata  & 46.02$\pm$0.01 \\
 XMM-LH &  168 & 1.958 & 46.95$\pm$0.01 & 8.85$\pm$0.23 &  \nodata & 3.74$\pm$0.06 &  \nodata &  \nodata & 42.49$\pm$0.14 &  \nodata \\
 XMM-LH &  176 & 1.533 & 45.77$\pm$0.01 & 8.12$\pm$0.04 & 3.43$\pm$0.01 &  \nodata & 3.55$\pm$0.01 & 43.19$\pm$0.02 &   \nodata & 45.30$\pm$0.01 \\
 XMM-LH &  191 & 0.787 & 45.94$\pm$0.01 & 7.94$\pm$0.02 & 3.86$\pm$0.01 &  \nodata & 3.82$\pm$0.18 & 42.53$\pm$0.02 &  \nodata  & 44.16$\pm$0.02 \\
 XMM-LH &  261 & 3.406 & 46.77$\pm$0.05 & 7.58$\pm$0.05 &  \nodata &  \nodata & 3.37$\pm$0.00 &  \nodata &  \nodata  & 45.02$\pm$0.02 \\
 XMM-LH &  270 & 1.576 & 45.66$\pm$0.01 & 8.39$\pm$0.06 & 3.75$\pm$0.01 & 3.55$\pm$0.05 & 3.73$\pm$0.14 & 43.50$\pm$0.02 & 41.93$\pm$0.11 & 45.17$\pm$0.01 \\
 XMM-LH &  321 & 1.008 & 45.04$\pm$0.03 & 7.44$\pm$0.03 &  \nodata &  \nodata & 3.55$\pm$0.01 & \nodata  &  \nodata  & 44.24$\pm$0.00 \\
 XMM-LH &  332 & 1.676 & 46.31$\pm$0.01 & 8.08$\pm$0.02 &  \nodata &  \nodata & 3.74$\pm$0.02 &  \nodata &  \nodata  & 44.64$\pm$0.01 \\
 XMM-LH &  354 & 3.409 & 46.95$\pm$0.02 & 8.93$\pm$0.17 &  \nodata &  \nodata & 3.76$\pm$0.16 &  \nodata &  \nodata  & 45.93$\pm$0.02 \\
 XMM-LH &  364 & 0.932 & 44.53$\pm$0.03 & 7.55$\pm$0.09 &   \nodata & \nodata & 3.65$\pm$0.04 & \nodata  &  \nodata  & 44.09$\pm$0.02 \\
 XMM-LH &  387 & 1.449 & 45.04$\pm$0.02 & 7.58$\pm$0.06 & 3.38$\pm$0.02 &  \nodata & 3.47$\pm$0.01 & 42.90$\pm$0.03 &   \nodata & 44.61$\pm$0.02 \\
 XMM-LH &  406 & 1.283 & 45.39$\pm$0.06 & 8.77$\pm$0.02 & 3.92$\pm$0.01 &  \nodata & 3.80$\pm$0.03 & 43.03$\pm$0.01 &  \nodata  & 44.86$\pm$0.01 \\
 XMM-LH &  430 & 1.553 & 46.48$\pm$0.02 & 8.33$\pm$0.17 &  \nodata &  \nodata & 3.90$\pm$0.17 &  \nodata &   \nodata & 44.54$\pm$0.02 \\
 XMM-LH &  453 & 1.214 & 45.02$\pm$0.06 & 7.41$\pm$0.04 &  \nodata &  \nodata & 3.52$\pm$0.02 &  \nodata &  \nodata  & 44.26$\pm$0.02 \\
 XMM-LH &  456 & 0.877 & 45.28$\pm$0.02 & 7.44$\pm$0.30 & 3.62$\pm$0.06 &  \nodata &  \nodata & 41.76$\pm$0.15 &  \nodata  &  \nodata \\
 XMM-LH &  475 & 1.205 & 47.07$\pm$0.01 & 8.90$\pm$0.00 & 3.69$\pm$0.00 &  \nodata & 3.87$\pm$0.14 & 44.13$\pm$0.00 &  \nodata  & 45.66$\pm$0.01 \\
 XMM-LH &  523 & 1.217 & 45.65$\pm$0.06 & 7.84$\pm$0.08 & 3.60$\pm$0.03 &  \nodata &  \nodata & 42.57$\pm$0.08 &  \nodata  & \nodata  \\
 XMM-LH &  529 & 1.940 & 46.74$\pm$0.04 & 8.55$\pm$0.11 &  \nodata & 3.68$\pm$0.09 &  \nodata &  \nodata & 42.16$\pm$0.20 &  \nodata \\
 XMM-LH &  532 & 1.675 & 45.01$\pm$0.03 & 8.29$\pm$0.31 & 3.60$\pm$0.03 &  \nodata & 3.73$\pm$0.02 & 42.91$\pm$0.06 &   \nodata & 45.01$\pm$0.01 \\
 XMM-LH &  553 & 1.440 & 46.36$\pm$0.01 & 8.68$\pm$0.03 & 3.76$\pm$0.01 &  \nodata &  \nodata & 43.49$\pm$0.02 &  \nodata  &  \nodata \\
 XMM-LH &  555 & 1.674 & 45.84$\pm$0.03 & 8.99$\pm$0.13 & 4.01$\pm$0.04 &  \nodata &  \nodata & 43.11$\pm$0.07 &  \nodata  &   \nodata \\
 XMM-LH &  591 & 1.535 & 45.65$\pm$0.05 & 8.22$\pm$0.08 & 3.62$\pm$0.32 &  \nodata &  \nodata & 43.18$\pm$0.00 &  \nodata  &   \nodata \\
 XMM-LH &  595 & 1.602 & 45.08$\pm$0.09 & 8.50$\pm$0.29 & 3.78$\pm$0.27 &  \nodata &  \nodata & 43.09$\pm$0.07 &  \nodata  &  \nodata \\
 XMM-LH &  604 & 2.104 & 46.47$\pm$0.02 & 8.66$\pm$0.83 & \nodata  & 3.54$\pm$0.62 &  \nodata &  \nodata & 42.86$\pm$0.00 &   \nodata \\
 XMM-LH & 2020 & 1.728 & 45.96$\pm$0.01 & 7.98$\pm$0.10 & 3.55$\pm$0.01 &  \nodata  &  \nodata & 42.99$\pm$0.02 & \nodata   &  \nodata \\
\end{longtable*}


\end{document}